\documentclass[11pt]{article}
\usepackage[utf8]{inputenc}
\usepackage{amsmath,amssymb}
\usepackage{graphicx}
\usepackage{subcaption}
\usepackage{natbib}
\usepackage{xcolor}
\usepackage{geometry}
\usepackage{hyperref}
\usepackage{booktabs}
\title{ Chaotic–Periodic Alternation in a Financial System with Double Delays}

\author{Jes\'us Salinas-Guti\'errez\thanks{Universidad Autónoma de la Ciudad de México} 
\and V\'{i}ctor F. Bre\~{n}a-Medina \thanks{Departamento de Matemáticas, Facultad de Ciencias. Universidad Nacional Autónoma de México}
\and Marcos \'Angel Gonz\'alez-Olvera\thanks{Universidad Aut\'onoma de la Ciudad de M\'exico, Colegio de Ciencia y Tecnolog\'ia, Mexico City, Mexico.} 
\and Anahí Flores-P\'erez\thanks{ Corresponding author. División de Ciencias Básicas, Facultad de Ingeniería. Universidad Nacional Autónoma de México} 
} 

\date{}

\begin{document}

\maketitle

\begin{abstract}
 
In this paper we present the chaotic–periodic alternation phenomenon within a three-dimensional nonlinear financial system subjected to Pyragas-type time-delayed feedback, as well as the lag effect of the interest rate in the price index dynamics. Through numerical analysis, transitions to chaos, including period-doubling cascades, period-halving routes, and incommensurable-period dynamics are characterized. The emergence of chaotic–periodic alternation are found and described. Results suggest that minor timing misalignments in the inner dynamics of a financial system, as well as those considered in the feedback loop can induce systemic volatility.
\end{abstract}

\section{Introduction}
\label{sec:intro}

Financial markets are known to be complex systems with highly irregular fluctuations that are not described purely by stochastic interpretations \citep{Kalecki1935,farmer2012complex}. In traditional approaches, periodic and aperiodic economic fluctuations are caused by stochastic exogenous shocks, but recent evidence suggests that nonlinear deterministic dynamics are capable of generating the inherent complexity observed in its dynamics originated in its structural nonlinearities \citep{Shone2002, AlvarezRamirez2001}.   %Therefore, market volatility may not merely be a reaction to external perturbations, but rather an endogenous manifestation of the system's structural nonlinearities.} 

Chaos theory offers paths to analyze the dynamics of financial systems, as it has demonstrated that in simple deterministic systems irregular and unpredictable trajectories can arise internally \citep{Klioutchnikov2017}. This has motivated the development of low-dimensional nonlinear models capable of capturing the essential dynamical features of financial systems, including periodic oscillations, quasi-periodicity, and deterministic chaos. By numerical bifurcation analysis, time delayed variables are shown to serve as bifurcation parameters, giving place to period-doubling cascades, attractor merging crises, and complex transitions between periodic and chaotic regimes. Such transitions mirror the chaotic–periodic alternation observed in other systems, explored previously in \citep{Flores2025}.

Time delays within the governing equations make possible to model economic decisions that depend on lagged information reception, that can come from regulatory processes, information asymmetries, and the finite speed at which market adjustments propagate. In this sense, the lag between the formulation of a policy decision and its implementation, and the additional lag between implementation and the observable effect on market variables \citep{Cesare2005, Fanti2007} have been described, resulting in delay differential equations (DDEs), which are infinite-dimensional in nature. 

Bifurcation structures that arise from these models with more complex dynamics than their ordinary differential equation counterparts have been reported. For example, in \citep{chen2008dynamics}, the dynamical evolution of a financial system is analyzed, specifically through the integration of the  Pyragas-type feedback loops \citep{Pyragas1992,Pyragas1995} applied to each state variable. Among those choices, it has been reported that a sensitive choice of delay magnitude and feedback intensity can lead to onset or suppression of chaos via delayed feedback control \citep{Holyst2000}.

The three-dimensional financial model proposed by Huang and Li \citep{HuangLi1993} and subsequently analyzed by Ma and Chen \citep{MaChen2001a, MaChen2001b}, describes the temporal evolution of the interest rate, investment demand, and price index as the main variables of a financial system, based on their coupled relations. This model considers the dependency between investment market contradictions, structural price adjustments, and the balance between supply and demand in commercial markets. Under certain parameter values, the system exhibits a chaotic attractor, making it a candidate benchmark for the study of financial instability \citep{zhang2014dynamical}. Extensions of this model have demonstrated that its chaotic properties are robust across a range of formulations, fractional-order generalizations \citep{ChenWC2008frac} and time-delay equations \citep{panigrahi2021stability,chen2008dynamics}. The role of heterogeneous delays acting concurrently on distinct state variables is still a area of research \citep{ccalics2022hopf,phukan2025bifurcation}.

In this work we examine the dynamics of the Ma-Chen financial system under a heterogeneous time-delayed feedback of information relying on the interest rate \citep{MaChen2001a, MaChen2001b}, with particular emphasis on the role of feedback polarity in determining the qualitative behaviour of the system. Two independent delays are associated with the interest rate and price index channels, respectively, while a feedback gain parametrizes the transition between cyclical and anti-cyclical policies given by external market regulators. To characterize the resulting dynamical responses, tools such as Bifurcation diagrams, Poincar\'{e} sections, phase portraits, and Lyapunov exponent estimates are employed. This extends the delayed feedback architecture established by Chen \citep{chen2008dynamics} and adapts it to describe the inner dynamics of a financial system when changes in the interest rate propagate with a lag. 

The main contributions of this work include the classification of the single-delay anti-cyclical dynamics of the Ma–Chen system, with findings of a period-halving route from chaos to period-1, as well as interleaved periodic and chaotic regimes associated with a second transmission delay. The analysis is done using a multi-diagnostic criterion for classifying regimes when the Lyapunov exponent is ambiguous.

The present paper is organized as follows: in Section 2 the financial model with delays is presented, as well as a description of the model. In Section 3 the model is analyzed and the computational framework is described. In Section 4 the results for different scenarios are obtained, where chaotic regimes, transitions and periodic regimes are identified; and discussion from these results is presented in Section 5. Finally, in Section 6 Conclusions are drawn.

\section{Problem Characterization}
\label{sec:model}

%\subsection{Baseline financial system}

Consider the nonlinear financial system proposed by \citep{HuangLi1993} and subsequently refined by \citep{MaChen2001a,MaChen2001b}. This model considers the coupled temporal dynamics of the interest rate $x(t)$, the investment demand $y(t)$, and the price index $z(t)$ expressed in its non-dimensionalized form as:
\begin{equation}
	\label{eq:baseline}
	\begin{cases}
		\dot{x} = z + (y - a)x\,, \\
		\dot{y} = 1 - by - x^2\,, \\
		\dot{z} = -x - cz\,,
	\end{cases}
\end{equation}
where $a > 0$ denotes the savings rate, $b > 0$ the cost per unit of investment, and $c > 0$ the elasticity of commercial demand. %{  The rate of change in $x$ is depends on market frictions and structural price re-alignments; the dynamics of $y$ are defined as a function of the investment rate, moderated by capital costs and prevailing interest rates; and the trajectory of $z$ characterizes the disequilibrium between aggregate supply and demand, subjected to inflationary pressure. It has been studied that \eqref{eq:baseline} shows a strange attractor and exhibits topological chaos under the distinguished parameter set $a = 3$, $b = 0.1$, and $c = 1$ \citep{MaChen2001b, chen2008dynamics}.}

%\subsection{Incorporation of heterogeneous time-delayed feedbacks}
%\subsection{ Heterogeneous time-delayed feedbacks distribution}

It has been reported that, in realistic financial environments, policy decisions and propagated market effects can show a lag \citep{borah2025stability}, so for this case two  distinct delay mechanisms are proposed. The first is the implementation delay $\tau_1$ following a Pyragas delayed feedback formulation \citep{Pyragas1995}, with a feedback term $K[x(t) - x(t - \tau_1)]$ in the interest rate equation, with $K \in \mathbb{R}$ is the response gain. The sign of $K$ carries an economic meaning: $K > 0$ corresponds to a \emph{cyclical} policy response, reinforcing the prevailing trend of $\dot{x}$ when comparing $x(t)$ and $x(t-\tau_1)$, whereas $K < 0$ corresponds to an \emph{anti-cyclical} response that opposes it. The second is the transmission delay $\tau_2$, representing the time required for interest rate adjustments to propagate into observable changes in the price index. %To describe the inherent asymmetries of the system, the assumption of delay commensurability is relaxed, which allows the temporal lags associated with distinct state-space components to independently evolve. %Such a heterogeneous formulation is crucial to acquire the non-synchronous coupling and disparate time scales that intrinsically characterize the system's dynamics.}
Thus, the extended system is given by

%{ In this fashion, the term $x(t) - x(t-\tau_1)$ corresponds to} the comparison of the interest rate at two { different times} separated by $\tau_1$, providing a measure of its rate of change over that horizon. The sign of $K$ carries economic meaning: $K > 0$ corresponds to a \emph{cyclical} policy response, reinforcing the prevailing trend, whereas $K < 0$ corresponds to an \emph{anti-cyclical} response, opposing it. The transmission delay $\tau_2$ enters structurally, replacing $x(t)$ by $x(t-\tau_2)$ in the price index equation, without an associated feedback gain. { Thus, we obtain the extended system}
%
\begin{gather}
	\label{eq:extended}
	\begin{cases}
		\dot{x}(t) = z(t) + \bigl[y(t) - a\bigr]x(t)
		             + K\bigl[x(t) - x(t-\tau_1)\bigr]\,, \\[4pt]
		\dot{y}(t) = 1 - by(t) - x(t)^2\,, \\[4pt]
		\dot{z}(t) = -x(t-\tau_2) - cz(t)\,.
	\end{cases}
\end{gather}

Notice that~\eqref{eq:extended} corresponds to a delay differential equation (DDE) system and is therefore infinite-dimensional, requiring specification of initial history functions $\phi_x : [-\tau_1, 0] \to \mathbb{R}$ and $\phi_z : [-\tau_2, 0] \to \mathbb{R}$ and a initial condition for  $y$ is a scalar $y_0 \in \mathbb{R}$.

\section{Equilibria Analysis and Numerical Methodology}
\label{sec:framework}

%{ This section provides a rigorous qualitative and quantitative characterisation of the dynamical regimes exhibited by system~\eqref{eq:extended}. Our methodology employs a complementary suite of numerical heuristics to map the system's behavioural landscape: bifurcation diagrams to identify critical parameter thresholds; Poincaré sections and phase-space portraits to resolve the topological structure of the emerging attractors; and the calculation of maximal Lyapunov exponents to provide a definitive measure of sensitive dependence on initial conditions.}

%\subsection{Constant solutions}

{From an analytical perspective, the stationary states of the extended system \eqref{eq:extended} coincide to those of the unperturbed baseline model~\eqref{eq:baseline}, as the Pyragas-type control law vanishes upon the system's convergence to an equilibrium point. Therefore, the x-coordinate of an equilibrium satisfies
\begin{gather}\label{eq:equimanifold}
	\xi x_*-x_*^3 = 0\,, \quad \textrm{where} \quad \xi = \dfrac{c-b-abc}{c}\,.
\end{gather}

The system thus admits a equilibrium $E_o$ and a pair of symmetric fixed points $E_{\pm}$, defined as
\begin{gather}\label{eq:equilibra}
	E_0 = \left(0,\dfrac{1}{b},0\right)\,, \quad E_{\pm} = \left(\pm\sqrt{\xi},\dfrac{1+ac}{c},\mp\dfrac{1}{c}\sqrt{\xi}\right)\,.
\end{gather}

The existence of $E_{\pm}$ depends on the parameter constraint $\xi > 0$. The subsequent numerical analysis is parameterized by the nominal vector $(a, b, c) = (3, 0.1, 1)$, based on the chaotic regime characterized in \citep{chen2008dynamics}. Given that the introduction of discrete delays requires a transition to an infinite-dimensional phase space, the system is initialized via history functions $\phi_i \in \mathcal{C}([-\tau_{\text{max}}, 0], \mathbb{R})$, where $\tau_{\text{max}}=\max\left\{\tau_1,\tau_2\right\}$. For this analysis, we consider a constant value for initial conditions $(x_0, y_0, z_0)$. Finally,  $\tau_1$, $\tau_2$ and $K$ are the parameters for bifurcation analysis.%, with the objective of resolving the stability boundaries and complex topological transitions emerging from the interaction of such a heterogeneous delay distribution.
}

\subsection{Computational framework and dynamical diagnostics}

The solutions for \eqref{eq:extended} are obtained using the \texttt{dde23} routine in \textsc{MATLAB} \citep{Shampine2001}, with a continuous-extension Runge--Kutta~(2,3) solver that tracks derivative discontinuities at delay breakpoints, with minimum step solution of $0.05$, with a relative tolerance of $10^{-5}$ and absolute tolerance of $10^{-7}$. Bifurcation diagrams are constructed from the integrated orbits on the Poincar\'e section $z=0$, $\dot z>0$, following the convention in \citep{chen2008dynamics}. In general, time is non-dimensional in the model, but at each parameter increment, a sufficiently long transient is discarded before recording, being typically 20 time units. Abrupt changes in the resulting point sets identify period-doubling cascades, saddle-node bifurcations, and attractor merging crises, the latter occurring when two coexisting chaotic attractors collide with an unstable periodic orbit on their shared basin boundary, producing a single enlarged attractor~\citep{Grebogi1983}, which is also reported for the baseline model~\citep{chen2008dynamics}.

These geometric observations are complemented by estimating the maximal Lyapunov exponent $\lambda_{\max}$. The variational method for ODEs~\citep{Benettin1980} has been extended to the infinite-dimensional DDE setting \cite{Farmer1982}, and $\lambda_{\max}$ is estimated, rather from the scalar time series $x(t)$ via phase-space reconstruction \citep{Wolf1985}, but made via a direct calculation of the euclidean distance between two perturbed solutions. The regime is classified as chaotic for $\lambda_{\max}>0$, marginally stable (quasi-periodic or at a bifurcation threshold) for $\lambda_{\max}=0$, and asymptotically stable for $\lambda_{\max}<0$.

Because time-series estimates of $\lambda_{\max}$ are susceptible to finite-data bias, two additional sources of spurious positive values must be considered. First, positive finite-time Lyapunov exponents (FTLEs) can appear during a transient preceding convergence to a stable limit cycle: larger transient FTLEs correspond to shorter escape times from the chaotic repeller, and smaller ones to longer transients~\citep{grantham1993, stefanski2010}. Second, the integration step size itself can induce ``computational chaos,'' yielding spuriously positive FTLEs for analytically periodic systems or, conversely, artificial periodicity in genuinely chaotic ones~\citep{demarkus2000}. For these reasons, a regime is classified as chaotic or periodic only when $\lambda_{\max}$, Poincar\'e mappings, phase portraits, and Fourier spectra jointly agree, following the multi-diagnostic approach used in \citep{Flores2025}. Ponicaré mappings are obtained for the plane $z=0$ with $x$ as the coordinate intersection in the positive direction.

\section{Numerical Results}
\label{sec:results}

The results are organized into groups of increasing complexity: single-delay anti-cyclical dynamics for $K<0$ in Section~\ref{subsec:anticyclic1}), and heterogeneous two-delay dynamics (Section~\ref{subsec:twodelay}).  Cases that reproduce the results of  \cite{chen2008dynamics}, considering $K>0$ for $\tau_1\neq 0$ are shown in Appendix \ref{Sec:ChenResults}. For each parameter configuration, the chaotic or non-chaotic nature of the attractor is established through the simultaneous assessment of four criteria: sensitive dependence on initial conditions, sign of $\lambda_{\max}$, periodicity or aperiodicity of the long-time solution, and the distribution of dominant frequencies in the Fourier spectrum.

Analysis cover the following scenarios for system \eqref{eq:extended}:
\begin{enumerate}
  %\item Single delay, with $K>0$, $\tau_1\neq 0$, $\tau_2=0$
  \item Single delay, with $K<0$, $\tau_1\neq 0$, $\tau_2=0$  
  \item Two delays, with constant $K>0$, $\tau_1$ and $\tau_2\neq0$
  \item Two delays, with constant $K<0$, $\tau_1$ and $\tau_2\neq0$  
\end{enumerate}

%  __                                
% (_   _   _ _|_ o  _  ._    |_|_ /| 
% __) (/_ (_  |_ | (_) | |     | o |  
%  

%---------------------------------------------------------------------
%\subsection{Single-delay cyclical dynamics ($K>0$, $\tau_2=0$)}
%\label{subsec:cyclic1}
%---------------------------------------------------------------------
% For this case,   $K=-0.1$ and $\tau_1=[1.5,4]$; and $K\in [-0.11,-0.085]$ and $\tau_1=1.85$ are analyzed.

%  __                                    
% (_   _   _ _|_ o  _  ._    |_|_ /|  /| 
% __) (/_ (_  |_ | (_) | |     | o | o | 
%                                        

%  __                              _  
% (_   _   _ _|_ o  _  ._    |_|_   ) 
% __) (/_ (_  |_ | (_) | |     | o /_ 
%                                     

%---------------------------------------------------------------------
\subsection{Single-delay anti-cyclical dynamics ($K<0$, $\tau_2=0$)}
\label{subsec:anticyclic1}
%---------------------------------------------------------------------

%  __                              _      
% (_   _   _ _|_ o  _  ._    |_|_   )  /| 
% __) (/_ (_  |_ | (_) | |     | o /_ o | 
%                                         

\subsubsection{Fixed $K=-0.1$, varying $\tau_1 \in [1.5, 4]$}%

{Considering an anti-cyclical policy while maintaining a single-delay architecture, $\tau_2 = 0$,  Figure~\ref{fig:FigPoinc2} shows the bifurcation diagram by varying delay range $1.5\leq\tau_1\leq 4$. In contrast to the classical supercritical cascade observed under cyclical feedback shown in Appendix \ref{Sec:ChenResults}, the anti-cyclical configuration promotes a period-halving route. As the temporal delay $\tau_1$ is progressively increased, the system transitions from chaos, to period-2 and period-1 regimes via successive period-halving bifurcations, re-entering into a chaotic state.

At the lower boundary of this parametric sweep, $\tau_1 = 1.69$, the system begins in a chaotic regime, confirmed by a positive maximal Lyapunov exponent, with sensitive dependence on initial conditions, and aperiodic time-series fluctuations and a broadband Fourier power spectrum, as shown in Fig. \ref{fig:FigSim_2}.b. As the delay changes into $\tau_1 \in [1.80, 1.95]$, the chaotic attractor disappears and the dynamics stabilize into a period-2 orbit, corroborated by two-loop continuous phase portraits, and verified by the existence of two dominant fundamental frequencies, and a negative maximal  Lyapunov exponent, as shown in Fig. \ref{fig:FigSim_2}.b-c.}

{In the interval $\tau_1 \in [1.95, 3.00]$, a subsequent period-halving bifurcation can be seen, recovering a stable, single-loop period-1 limit cycle. Within these bounds, the system returns to a periodic regime. As the delay reaches $\tau_1 = 3.05$, the period-1 orbit changes into a chaotic regime, with a transition to the sequence of period-2, period-1 and chaos. This can be interpreted as vulnerability in delayed feedback control, where policies implemented with excessive temporal lag do not stabilize the market, but actively can give rise to an erratic behavior.

%  _                    _  
% |_ o  _      ._ _    (_) 
% |  | (_| |_| | (/_   (_) 
%       _|                 
%FIGURA - Bifurcation diagram - tau1 = 1.69 y tau1 = 3.05
\begin{figure}[t!]
	\centering
	\begin{subfigure}[b]{0.75\textwidth}
		\includegraphics[width=\linewidth]{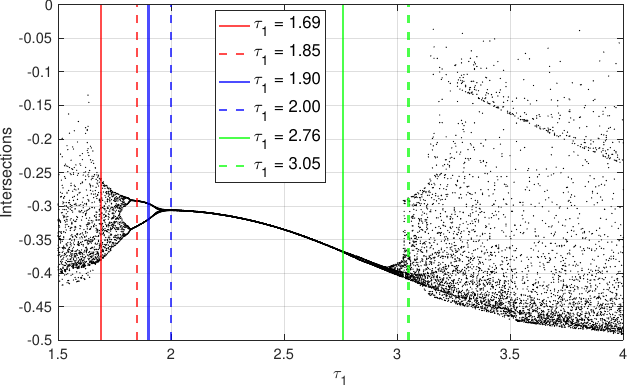}
		\caption{Bifurcation diagram for $\tau_1$ with crossings of $z=0$.}\label{fig:FigPoinc2}
	\end{subfigure}\\
	\begin{subfigure}[b]{0.95\textwidth}
		\includegraphics[width=\linewidth]{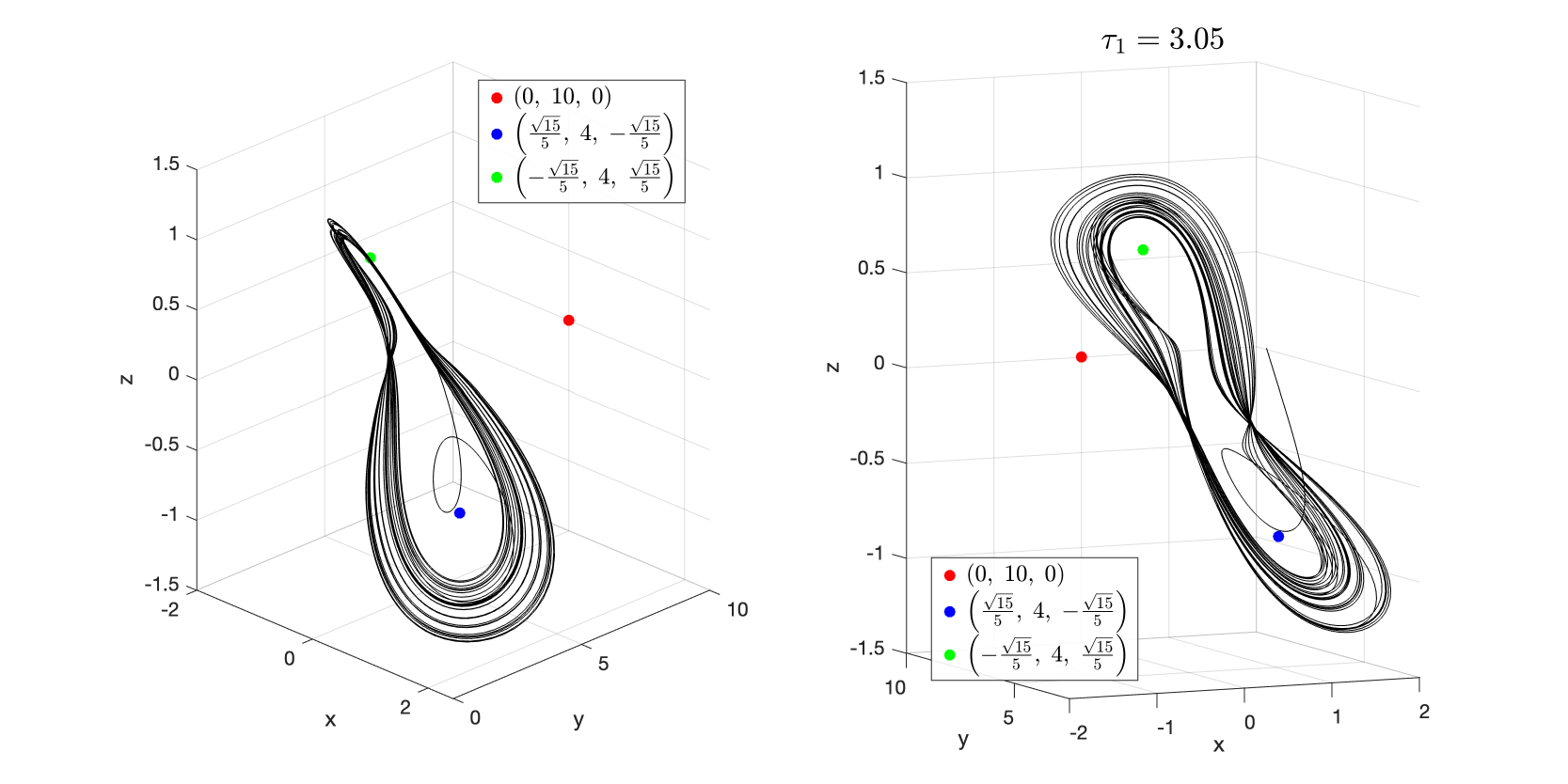}%Atractor_3(1).png y Atractor_3(2).png
		\caption{Phase portraits for $\tau_1 = 1.69$ and $\tau_1 = 3.05$}
	\end{subfigure}
	\caption{ (a)~Bifurcation diagram on the Poincar\'{e} section $z = 0$. A period-halving route chaos $\to$ period-2 $\to$ period-1 $\to$ chaos is observed; (b) Phase portraits at the two selected values of  $\tau_1$ that show chaotic behaviour for an anti-cyclical action. Parameter set values: $K = -0.1$ and $\tau_2 = 0$.}
	\label{fig:Atractores_3}
\end{figure}

%  _                   __ 
% |_ o  _      ._ _     / 
% |  | (_| |_| | (/_   /  
%  
\begin{figure}[t!]
	\centering
	\begin{subfigure}[b]{0.48\textwidth}
		\includegraphics[width=\linewidth]{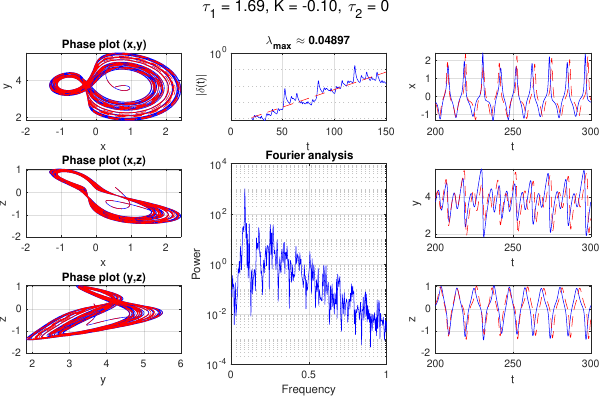}%FigSim_1(2).pdf
		\caption{$\tau_1 = 1.69$ (chaos)}
	\end{subfigure}\hfill
	\begin{subfigure}[b]{0.48\textwidth}
		\includegraphics[width=\linewidth]{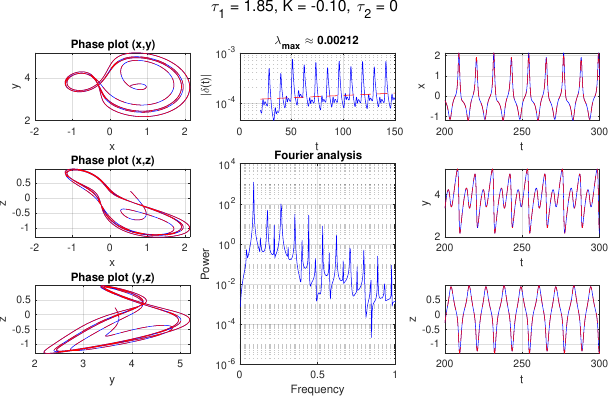}%FigSim_2(2).pdf
		\caption{$\tau_1 = 1.85$ (period-2)}
	\end{subfigure}
	\vspace{0.3cm}
	\begin{subfigure}[b]{0.48\textwidth}
		\includegraphics[width=\linewidth]{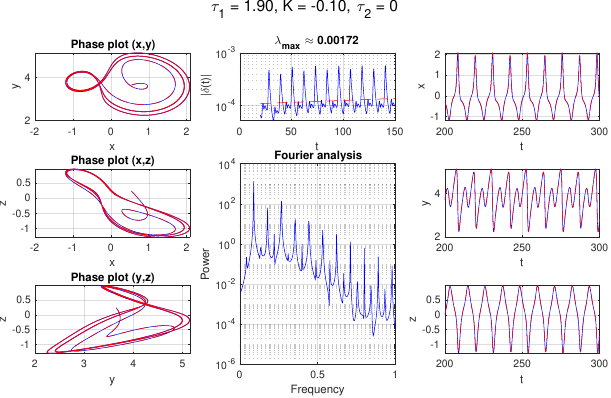}%FigSim_3(2).pdf
		\caption{$\tau_1 = 1.90$ (period-2)}
	\end{subfigure}\hfill
	\begin{subfigure}[b]{0.48\textwidth}
		\includegraphics[width=\linewidth]{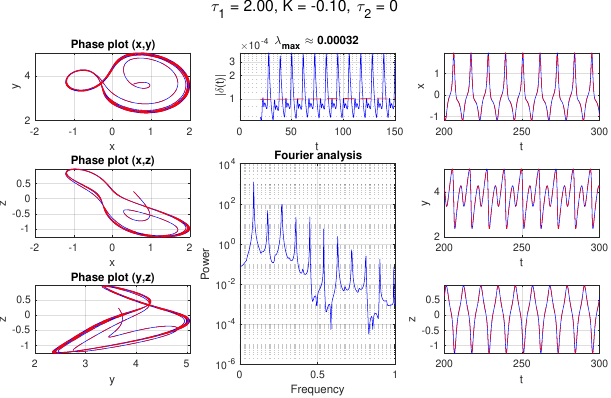}%FigSim_4(2).pdf
		\caption{$\tau_1 = 2.00$ (period-1)}
	\end{subfigure}
	\vspace{0.3cm}
	\begin{subfigure}[b]{0.48\textwidth}
		\includegraphics[width=\linewidth]{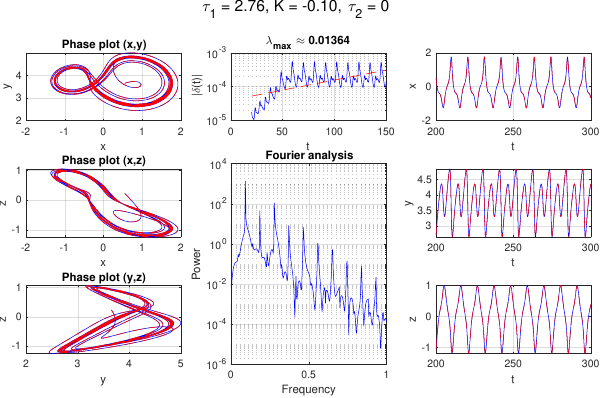}%FigSim_5(2).pdf
		\caption{$\tau_1 = 2.76$ (period-1)}
	\end{subfigure}\hfill
	\begin{subfigure}[b]{0.48\textwidth}
		\includegraphics[width=\linewidth]{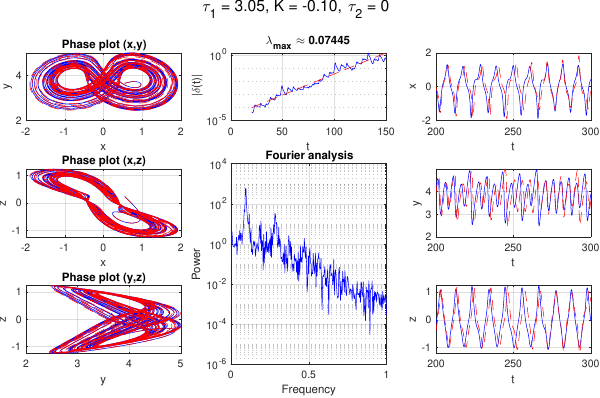}%FigSim_6(2).pdf
		\caption{$\tau_1 = 3.05$ (chaos)}
	\end{subfigure}
	\caption{Phase portraits for $K = -0.1$, $\tau_2 = 0$, and selected values of
		$\tau_1$, illustrating the  period-halving route and chaotic behaviour.}
	\label{fig:FigSim_2}
\end{figure}

% ___                _  
%  |  _. |_  |  _    _) 
%  | (_| |_) | (/_   _) 
%                       
\begin{table}[t!]	 
	\centering
	\caption{Dynamical regimes for $K = -0.1$, $\tau_2 = 0$, varying $\tau_1$.}
	\label{tabla_3}
	\begin{tabular}{r|cccccc}
		\toprule
		$\tau_1$ & 1.69 & 1.85 & 1.90 & 2.00 & 2.76 & 3.05 \\ 
		$\lambda_{\max}$ & $> 0$ & $\approx 0$ & $\approx 0$ & $\approx 0$ & $>0$ & $> 0$\\
		Dynamics & Chaos & Period-2 & Period-2 & Period-1 & Period-1 & Chaos\\
		\bottomrule
	\end{tabular}
\end{table}

The distinct strange attractors bounding this stable window, from $\tau_1 = 1.69$ to $\tau_1 = 3.05$, are shown in Figure~\ref{fig:Atractores_3}. It can be seen that the terminal chaotic manifold exhibits a higher degree of phase-space stretching and folding, alongside a significantly larger amplitude, when compared to its initial counterpart. This geometric expansion translates to fluctuations in the dynamics, and those changes from limit cycles back to strange attractors are shown in Figure~\ref{fig:FigSim_2}. Furthermore, the corresponding spectral metrics, including the transition of the maximal Lyapunov exponent from strictly negative back to positive values, can be seen in Table~\ref{tabla_3}.}

%  __                              _    _  
% (_   _   _ _|_ o  _  ._    |_|_   )    ) 
% __) (/_ (_  |_ | (_) | |     | o /_ o /_ 
%                                          

\subsubsection{Fixed $\tau_1 = 1.85$, varying $K \in [-0.11, -0.085]$}

{Initiating from the period-2 configuration with the parameter values $\tau_1=1.85$ and $K =  -0.11$, the anti-cyclical feedback gain $K$ is subjected to a variation from this value up to $K=-0.085$, while maintaining a single-delay ($\tau_2 = 0$). The resulting bifurcation diagram, illustrated in Figure~\ref{fig:FigPoinc4}, captures a perioud-doubling cascade to chaos. Specifically, as the magnitude of the feedback intensity is gradually attenuated, the dynamics undergoes a sequence of period-doubling bifurcations, transitioning from a stable period-1 limit cycle when $-0.11 \leq K < -0.10$, into a period-2 orbit for $-0.10 \leq K < -0.0985$, subsequently splitting into a period-4 configuration for $-0.0985 \leq K < -0.09$, and ultimately accumulating into a chaotic regime at $K = -0.085$.

%  _                    _  
% |_ o  _      ._ _    (_| 
% |  | (_| |_| | (/_     | 
%       _|                 
\begin{figure}[t]
	\centering
	\includegraphics[width=0.55\linewidth]{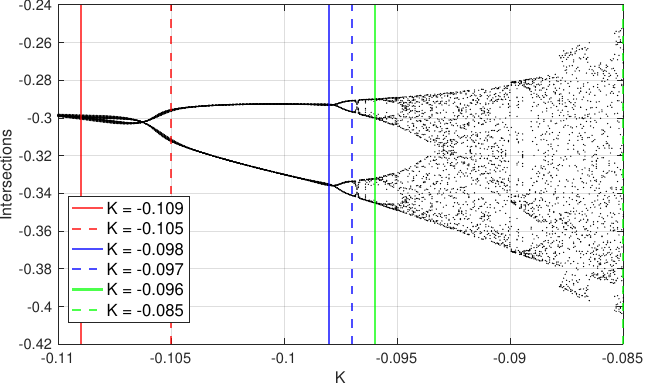}%FigPoinc4_B.pdf
	\caption{Bifurcation diagram on the Poincar\'{e} section $z = 0$ as a function
		of $K$, with $\tau_1 = 1.85$ and $\tau_2 = 0$. A period-doubling cascade
		period-1 $\to$ period-2 $\to$ period-4 $\to$ chaos is observed.}
	\label{fig:FigPoinc4}
\end{figure}

\begin{table}[t!]
	 
	\centering
	\caption{Dynamical regimes for $\tau_1 = 1.85$, $\tau_2 = 0$, varying $K<0$.}
	\label{tabla_4}
	\begin{tabular}{r|cccccc}
		\toprule
		$K$ & $-0.109$ & $-0.105$ & $-0.098$ & $-0.097$ & $-0.096$ & $-0.085$ \\ 
		$\lambda_{\max}$ & $\approx 0$ & $< 0$ & $\approx 0$ & $> 0$ & $> 0$ & $> 0$\\
		Dynamics & Period-1 & Period-2 & Period-2 & Period-4 & Period-4 & Chaos\\
		\bottomrule
	\end{tabular}
\end{table}

%  _                       _  
% |_ o  _      ._ _    /| / \ 
% |  | (_| |_| | (/_    | \_/ 
%       _|                    
\begin{figure}[t!]
	\centering
	\includegraphics[width=0.75\linewidth]{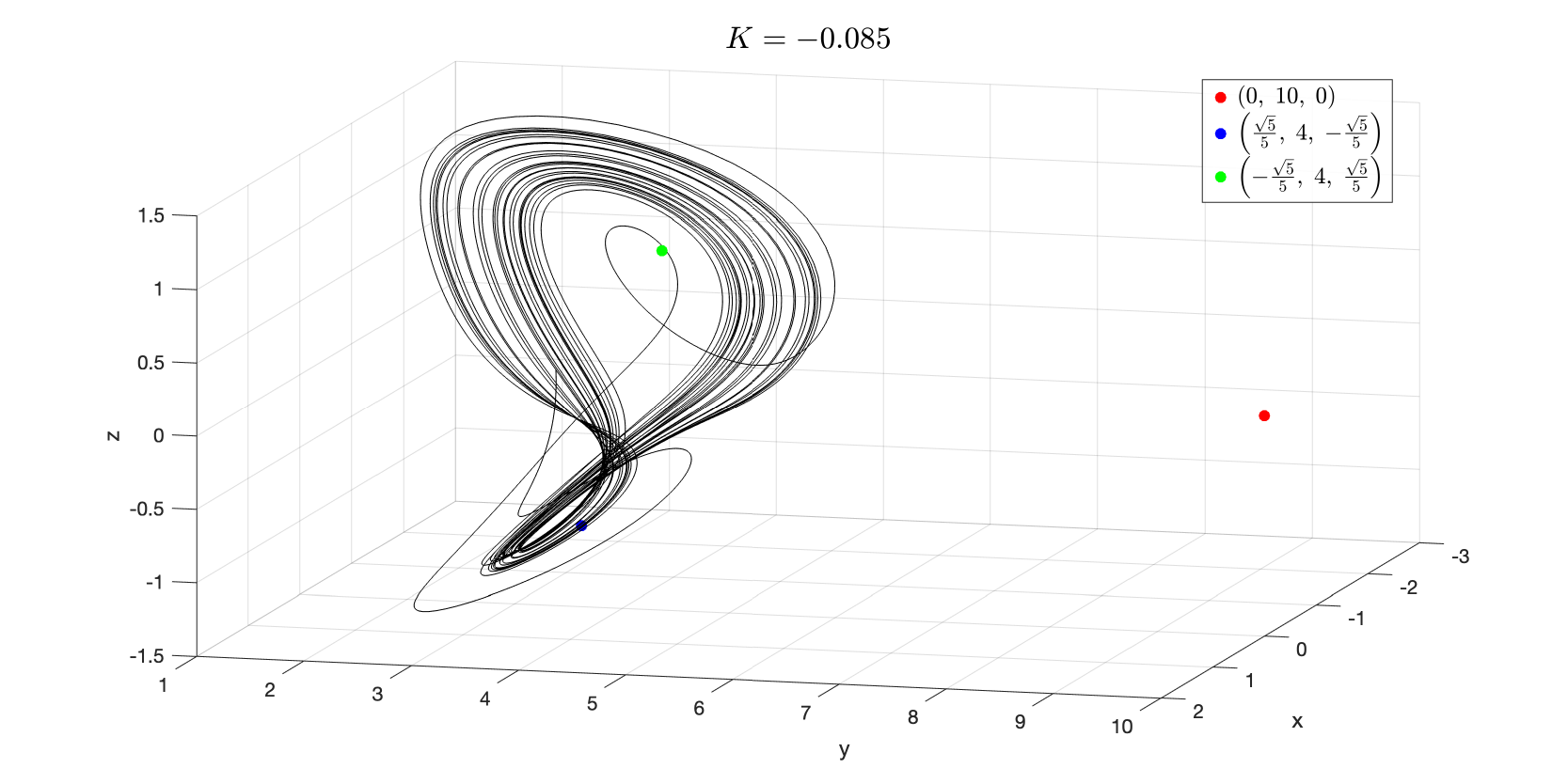}%Atractor_4(1).png
	\caption{Strange attractor for $\tau_1 = 1.85$, $\tau_2 = 0$, $K = -0.085$.}
	\label{fig:Atractor_4}
\end{figure}

%  _                         
% |_ o  _      ._ _    /| /| 
% |  | (_| |_| | (/_    |  | 
%       _|                   
\begin{figure}[t!]
	\centering
	\begin{subfigure}[b]{0.48\textwidth}
		\includegraphics[width=\linewidth]{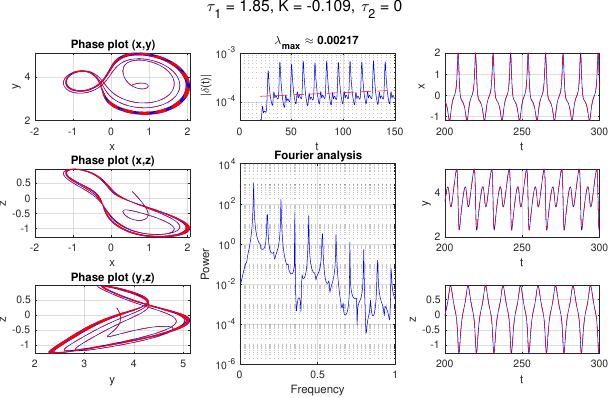}%FigSim_1(4).pdf
		\caption{$K = -0.109$ (period-1)}
	\end{subfigure}\hfill
	\begin{subfigure}[b]{0.48\textwidth}
		\includegraphics[width=\linewidth]{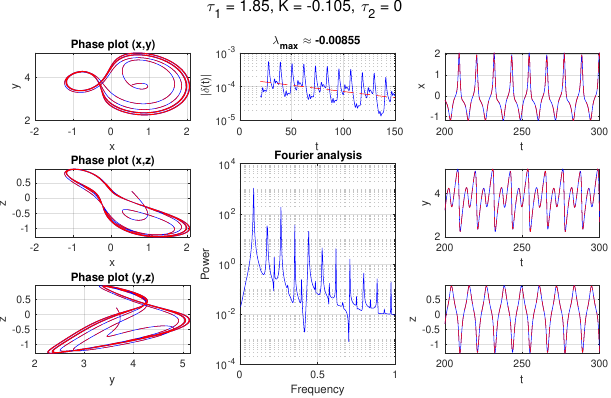}%FigSim_2(4).pdf
		\caption{$K = -0.105$ (period-2)}
	\end{subfigure}
	\vspace{0.3cm}
	\begin{subfigure}[b]{0.48\textwidth}
		\includegraphics[width=\linewidth]{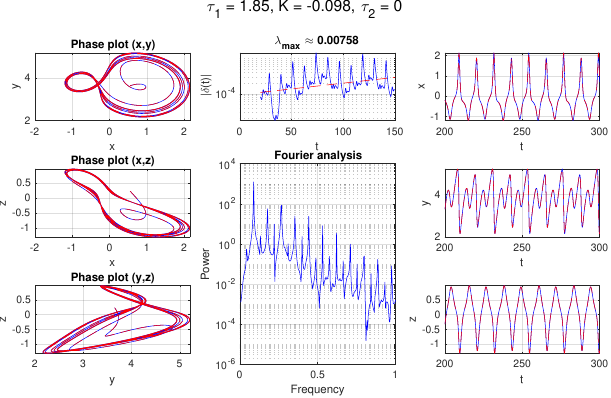}%FigSim_3(4).pdf
		\caption{$K = -0.098$ (period-2)}
	\end{subfigure}\hfill
	\begin{subfigure}[b]{0.48\textwidth}
		\includegraphics[width=\linewidth]{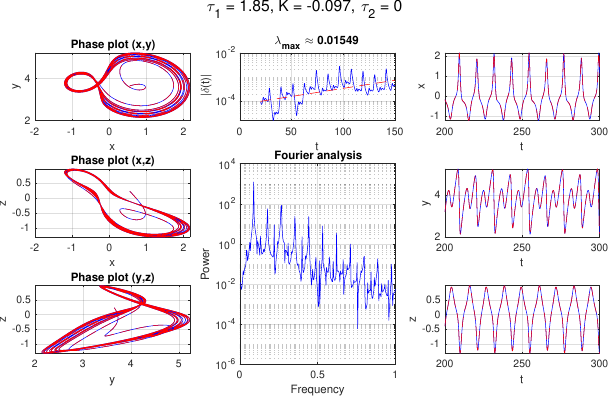}%FigSim_4(4).pdf
		\caption{$K = -0.097$ (period-4)}
	\end{subfigure}
	\vspace{0.3cm}
	\begin{subfigure}[b]{0.48\textwidth}
		\includegraphics[width=\linewidth]{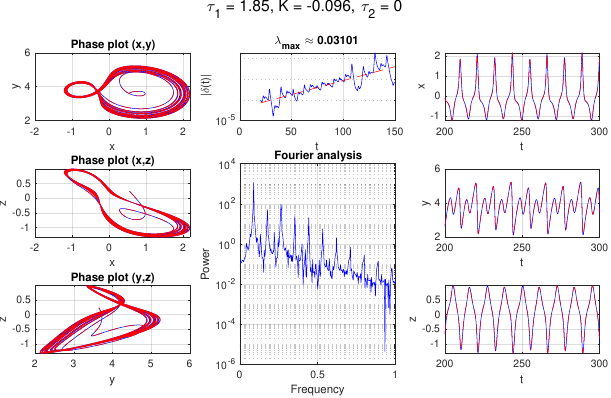}%FigSim_5(4).pdf
		\caption{$K = -0.096$ (period-4)}
	\end{subfigure}\hfill
	\begin{subfigure}[b]{0.48\textwidth}
		\includegraphics[width=\linewidth]{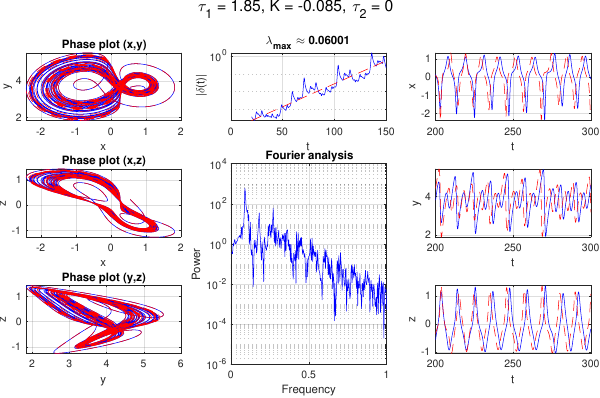}%FigSim_6(4).pdf
		\caption{$K = -0.085$ (chaos)}
	\end{subfigure}
	\caption{Phase portraits for $\tau_1 = 1.85$, $\tau_2 = 0$, and selected values
		of $K < 0$, illustrating the period-doubling cascade to chaos.}
	\label{fig:FigSim_4}
\end{figure}

Across the period-1, period-2, and period-4 windows, time-series estimations of the maximal Lyapunov exponent inconsistent results on positivity on periodic dynamics; however, geometric and spectral cross-validation confirms the absence of the  dependence on initial conditions expected in chaotic regimes, as trajectories remaining strictly confined to regular, repeating limit cycles. This discrepancy shows that estimates of $\lambda_{max}$ are insufficient to characterize chaotic regimes, and more numerical tools are needed to confirm it.

Deterministic chaos is found at $K = -0.085$, were the aforementioned multi-diagnostic criteria (sensitive dependence, erratic and aperiodic state trajectories, and a continuous, broadband Fourier spectral density) are met. The trajectories of the resulting strange attractor are shown in Figure~\ref{fig:Atractor_4}, while phase portraits are shown in Figure~\ref{fig:FigSim_4}., with the corresponding metrics in Table~\ref{tabla_4}.}

%  __                              _  
% (_   _   _ _|_ o  _  ._    |_|_  _) 
% __) (/_ (_  |_ | (_) | |     | o _) 
%                                     

%---------------------------------------------------------------------
\subsection{Two-delay dynamics}
\label{subsec:twodelay}
%---------------------------------------------------------------------
In this case, analysis for $K=0.35$, $\tau_1=6.40$, $\tau_2\in[0.9,1.3]$; and $K=-0.098$, $\tau_1=1.85$, $\tau_2\in[0.5,0.7]$ are presented. In order to isolate and characterize the impact of this secondary lag, we perturb two dynamically distinct, non-chaotic baseline configurations previously established in the analyses of Sections~\ref{subsec:cyclic1} and~\ref{subsec:anticyclic1}.

\subsubsection{Cyclical case: $\tau_1 = 6.40$, $K = 0.35$, varying
	$\tau_2 \in [0.9, 1.3]$}

%The first analysis begins at $\tau_1=6.40$ and $K = 0.35$, where the system operates within an incommensurable-period regime characterized by dense quasi-periodic motion on a two-dimensional invariant torus. The second baseline, situated within the anti-cyclical feedback domain at $\tau_1=1.85$ and $K = -0.098$, exhibits a robust subharmonic period-2 limit cycle. Upon introducing the transmission delay into these specific, pre-stabilized setting, we aim to expose the complex combinatorial interplay between the implementation and transmission channels. This methodological approach allows us to investigate how asymmetric delay couplings interact to induce intermittent chaotic bursting, and fundamentally alter the dynamics of the macroeconomic system.

{ Figure~\ref{fig:FigPoinc5} shows the bifurcation diagram as a function of $\tau_2 \in [0.9, 1.3]$, as can be seen in panel~(a), with panel~(b) providing a magnified resolution of the critical sub-domain $\tau_2 \in [1.02, 1.04]$. It can be seen how there are abrupt transitions between ordered limit sets and chaotic attractors. This switching serves as an indication of chaotic–periodic alternation, which has been extensively described in recent literature \citep{Flores2025,intermitencia}.

\begin{figure}[t!]
	\centering
	\begin{subfigure}[b]{0.48\textwidth}
		\includegraphics[width=\linewidth]{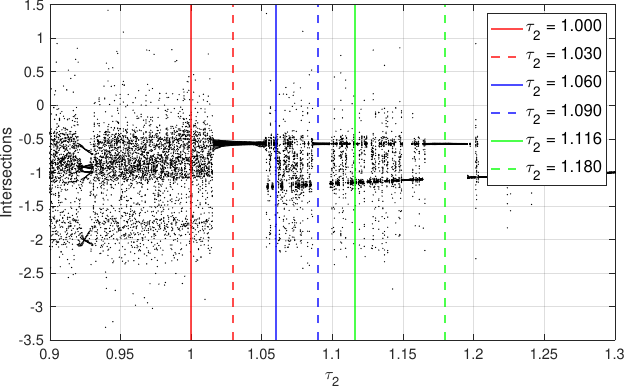}%FigPoinc5_B.pdf
		\caption{$\tau_2 \in [0.9, 1.3]$}
	\end{subfigure}\hfill
	\begin{subfigure}[b]{0.48\textwidth}
		\includegraphics[width=\linewidth]{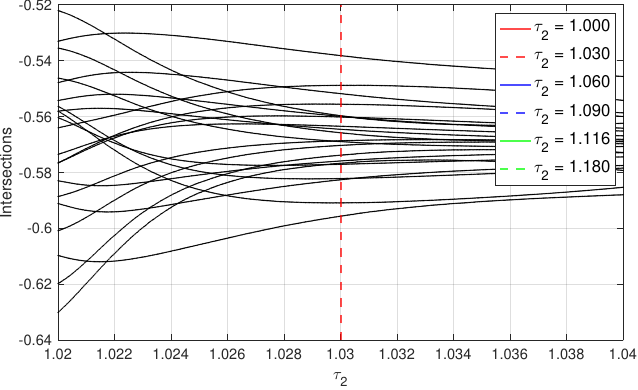}%Refinamiento_1.pdf
		\caption{Refined view: $\tau_2 \in [1.02, 1.04]$}
	\end{subfigure}
	\caption{Bifurcation diagram on the Poincar\'{e} section $z = 0$ as a function
		of $\tau_2$, with $\tau_1 = 6.40$ and $K = 0.35$. Recurrent transitions
		between chaotic and periodic regimes over small $\tau_2$ increments are
		characteristic of chaotic–periodic alternation as is shown in panel~(a). Panel~(b) resolves the fine
		structure near $\tau_2 \approx 1.03$.}
	\label{fig:FigPoinc5}
\end{figure}

\begin{table}[t!]
	 
	\centering
	\caption{Dynamical regimes for $\tau_1 = 6.40$, $K = 0.35$, varying $\tau_2$.}
	\label{tabla_5}
	\begin{tabular}{r|cccccc}
		\toprule
		$\tau_2$ & $1.00$ & $1.03$ & $1.06$ & $1.09$ & $1.116$ & $1.18$ \\ 
		$\lambda_{\max}$ & $> 0$ & $\approx 0$ & $> 0$ & $\approx 0$ & $> 0$ & $\approx 0$\\%0.041917
		Dynamics & Chaos & Incommensurable period & Chaos & Period-1 & Chaos & Period-1\\
		\bottomrule
	\end{tabular}
\end{table}

\begin{figure}[t!]
	\centering
	\begin{subfigure}[b]{0.48\textwidth}
		\includegraphics[width=\linewidth]{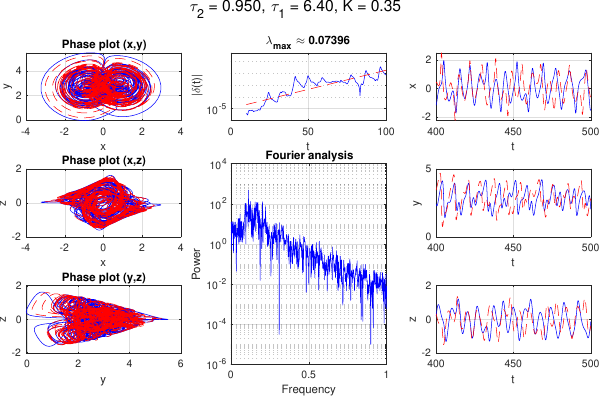}%FigSim_1(5).pdf
		\caption{$\tau_2 = 0.95$ (chaos, $\lambda_{\max}=0.07396$)}
	\end{subfigure}\hfill
	\begin{subfigure}[b]{0.48\textwidth}
		\includegraphics[width=\linewidth]{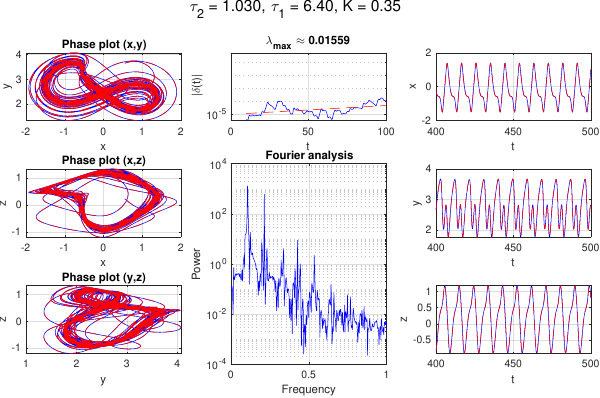}%FigSim_2(5).pdf
		\caption{$\tau_2 = 1.03$ (incommensurable period)}
	\end{subfigure}
	\vspace{0.3cm}
	\begin{subfigure}[b]{0.48\textwidth}
		\includegraphics[width=\linewidth]{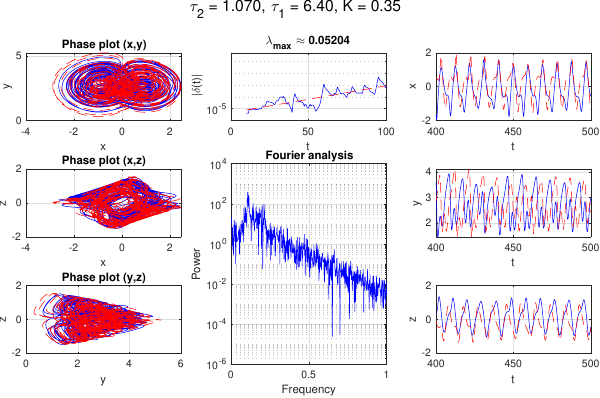}%FigSim_3(5).pdf
		\caption{$\tau_2 = 1.07$ (chaos)}
	\end{subfigure}\hfill
	\begin{subfigure}[b]{0.48\textwidth}
		\includegraphics[width=\linewidth]{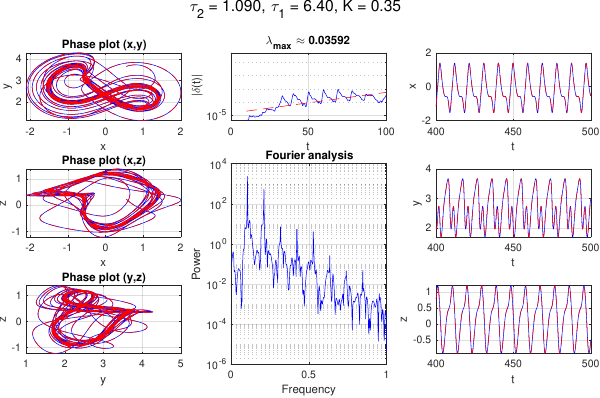}%FigSim_4(5).pdf
		\caption{$\tau_2 = 1.09$ (period-1)}
	\end{subfigure}
	\vspace{0.3cm}
	\begin{subfigure}[b]{0.48\textwidth}
		\includegraphics[width=\linewidth]{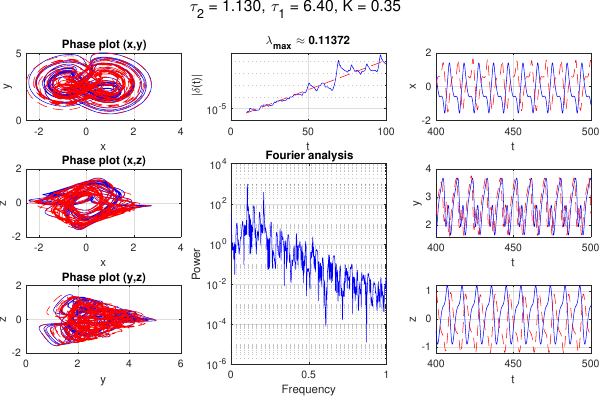}%FigSim_5(5).pdf
		\caption{$\tau_2 = 1.13$ (chaos)}
	\end{subfigure}\hfill
	\begin{subfigure}[b]{0.48\textwidth}
		\includegraphics[width=\linewidth]{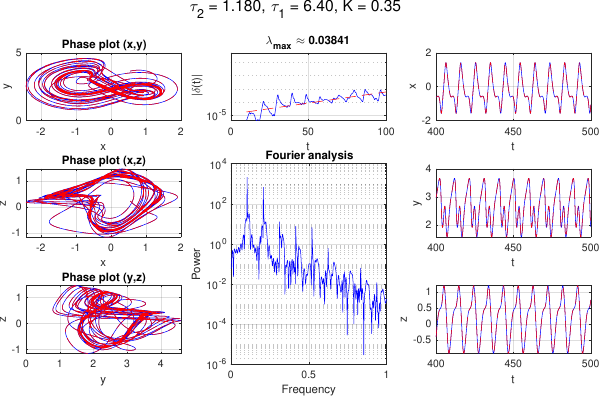}%FigSim_6(5).pdf
		\caption{$\tau_2 = 1.18$ (period-1)}
	\end{subfigure}
	\caption{Phase portraits for $K = 0.35$, $\tau_1 = 6.40$, and selected values
		of $\tau_2$, illustrating chaotic–periodic alternation.}
	\label{fig:FigSim_5}
\end{figure}

\begin{figure}[t!]
	\centering
	\includegraphics[width=0.85\textwidth]{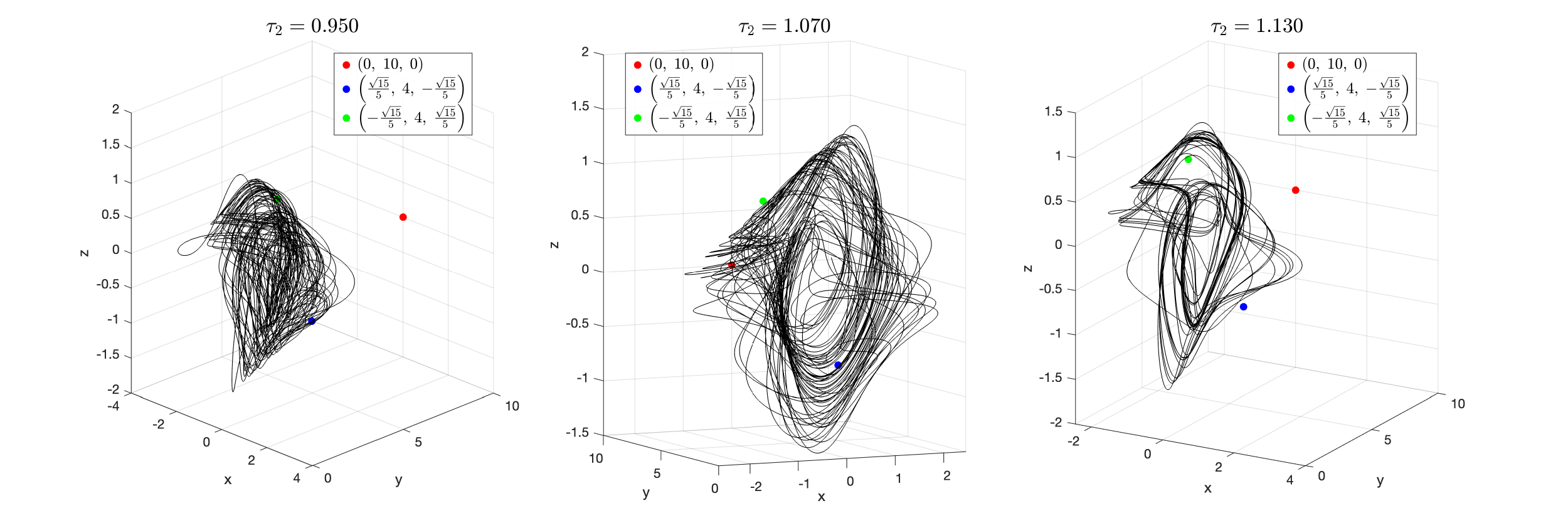}
%	\begin{subfigure}[b]{0.32\textwidth}
%		\includegraphics[height=4.5cm,keepaspectratio]{images/Atractor_5(1)}
%		\caption{$\tau_2 = 1.00$}
%	\end{subfigure}\hfill
%	\begin{subfigure}[b]{0.32\textwidth}
%		\includegraphics[height=4.5cm,keepaspectratio]{images/Atractor_5(2)}
%		\caption{$\tau_2 = 1.06$}
%	\end{subfigure}\hfill
%	\begin{subfigure}[b]{0.32\textwidth}
%		\includegraphics[height=4.5cm,keepaspectratio]{images/Atractor_5(3)}
%		\caption{$\tau_2 = 1.116$}
%	\end{subfigure}
	\caption{Strange attractors for $\tau_1 = 6.40$, $K = 0.35$, at the three
		chaotic values of $\tau_2$ identified in Figure~\ref{fig:FigPoinc5}.}
	\label{fig:Atractores_5}
\end{figure}

At $\tau_2 = 1.00$, the system is chaotic, as verified by a positive maximal Lyapunov exponent of $\lambda_{\max} = 0.041917$, highly aperiodic state trajectories, and a  broadband Fourier spectral density seen in some nearby values in Fig.\ref{fig:FigSim_5}. However, a minimal parametric change to $\tau_2 = 1.03$ forces a sudden change onto an incommensurable-period torus where exponential trajectory divergence is lost ($\lambda_{\max} \approx 0$). This quasi-periodic stability proves highly transient; as the delay is pushed into the window $1.05<\tau_2<1.08$, where it breaks into chaos ($\lambda_{\max} > 0$), and then settles a change at $1.08\leq\tau_2\leq1.10$ into a stable, single-loop period-1 limit cycle, confirmed by a Lyapunov negative exponent. This persists as the parameter increases, and chaos re-emerges across $1.10\leq\tau_2\leq1.15$, to be suppressed by a restored period-1 orbit for $1.15\leq\tau_2 \leq1.20$. The distinct strange attractors that show these unstable windows, specifically at $\tau_2 = 1.00,1.06$, and $1.116$, are shown in Figure~\ref{fig:Atractores_5}. Selected parameter selections and metrics are shown in Figure~\ref{fig:FigSim_5} and Table~\ref{tabla_5}.

The alternation between deterministic chaos and periodic orbits shows that the transmission delay $\tau_2$ has a key role in the dynamics. Even when the isolated implementation lag $\tau_1$ and cyclical feedback gain $K$ are calibrated to obtain a stable quasi-periodic motion, the delay in the information propagation of $x$ into the dynamics of $z$ can induce chaotic bursts. From a macroeconomic perspective, this can model a design vulnerability, as even if stabilization policies are designed with a Pyragas-type feedback, un-modeled transmission delays can keep the periodic regime, but in other cases give rise to chaotic dynamics if not considered.}

\subsubsection{Anti-cyclical case: $\tau_1 = 1.85$, $K = -0.098$, varying
	$\tau_2 \in [0.5, 0.7]$}

\begin{figure}[t!]
	\centering
	\begin{subfigure}[b]{0.48\textwidth}
		\includegraphics[width=\linewidth]{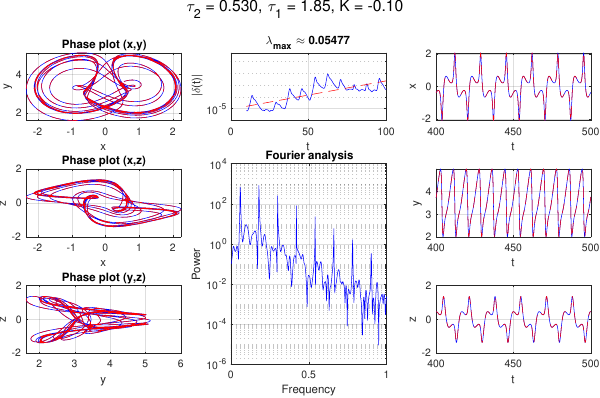}%FigSim_1(6).pdf
		\caption{$\tau_2 = 0.53$ (period-1)}
	\end{subfigure}\hfill
	\begin{subfigure}[b]{0.48\textwidth}
		\includegraphics[width=\linewidth]{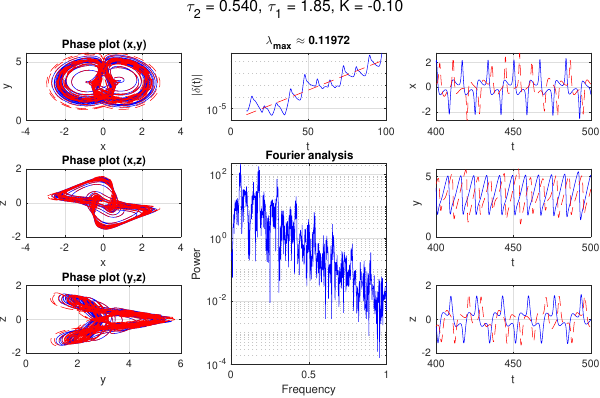}%FigSim_2(6).pdf
		\caption{$\tau_2 = 0.54$ (chaos)}
	\end{subfigure}
	\vspace{0.3cm}
	\begin{subfigure}[b]{0.48\textwidth}
		\includegraphics[width=\linewidth]{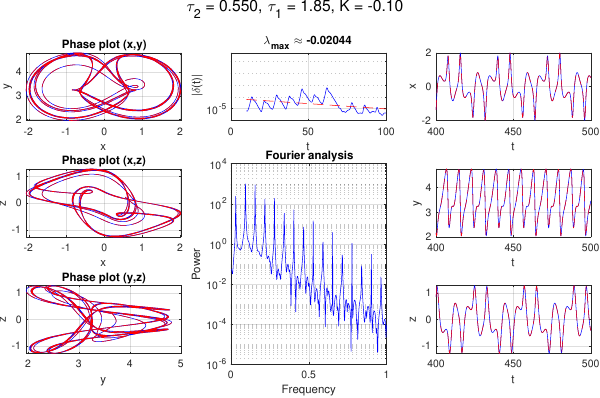}%FigSim_3(6).pdf
		\caption{$\tau_2 = 0.55$ (period-2)}
	\end{subfigure}\hfill
	\begin{subfigure}[b]{0.48\textwidth}
		\includegraphics[width=\linewidth]{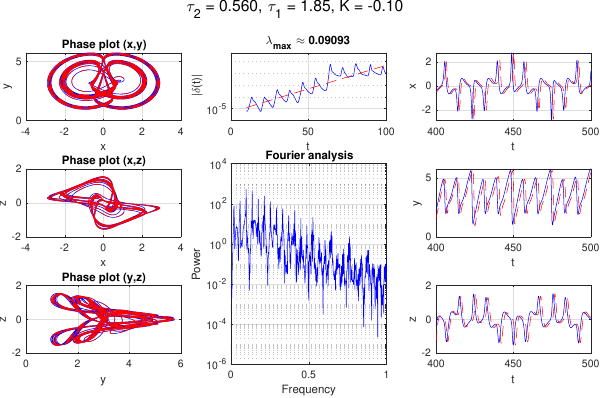}%FigSim_4(6).pdf
		\caption{$\tau_2 = 0.56$ (chaos)}
	\end{subfigure}
	\vspace{0.3cm}
	\begin{subfigure}[b]{0.48\textwidth}
		\includegraphics[width=\linewidth]{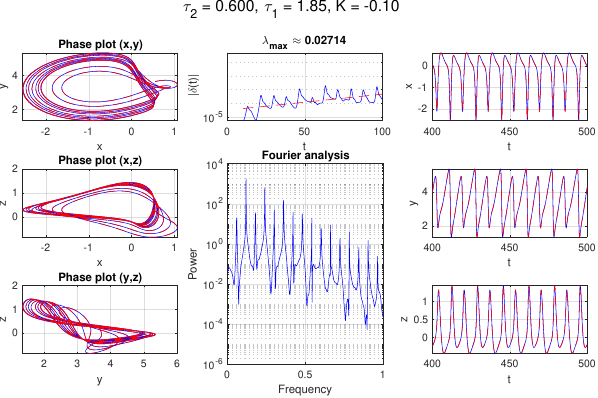}%FigSim_5(6).pdf
		\caption{$\tau_2 = 0.60$ (period-2)}
	\end{subfigure}\hfill
	\begin{subfigure}[b]{0.48\textwidth}
		\includegraphics[width=\linewidth]{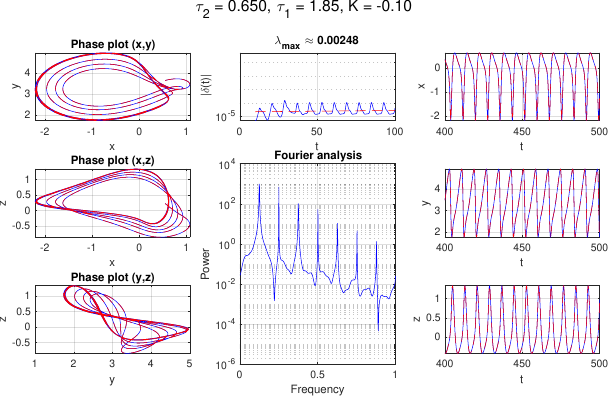}%FigSim_6(6).pdf
		\caption{$\tau_2 = 0.65$ (period-1)}
	\end{subfigure}
	\caption{Phase portraits for $K = -0.098$, $\tau_1 = 1.85$, and selected values
		of $\tau_2$. Panels (a)--(d): chaotic–periodic alternation. Panels (e)--(f):
		period-halving route.}
	\label{fig:FigSim_6}
\end{figure}

{ Figure~\ref{fig:FigPoinc6} shows the Poincaré section over $0.5 \leq \tau_2 \leq 0.7$, while holding the anti-cyclical baseline at $\tau_1 = 1.85$ and $K = -0.098$. The resulting dynamics depicts the sequential unfolding of two distinct phenomena: a localized regime of chaos spanning $0.53\leq\tau_2 \leq0.59$, followed by a stabilizing period-halving sequence for $\tau_2 > 0.59$.

\begin{figure}[t!]
	\centering
	\includegraphics[width=0.55\linewidth]{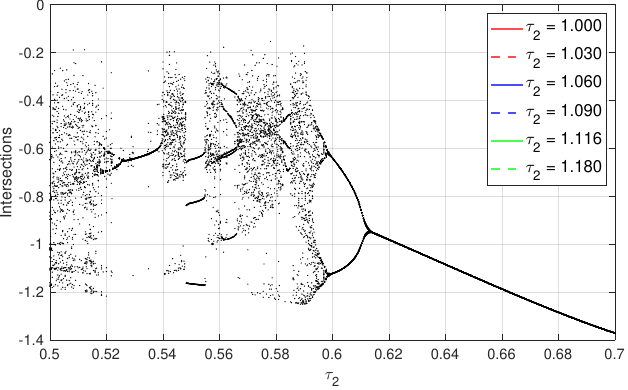}%FigPoinc6_B.pdf
	\caption{Bifurcation diagram on the Poincar\'{e} section $z = 0$ as a function
		of $\tau_2$, with $\tau_1 = 1.85$ and $K = -0.098$. Chaotic–periodic alternation for
		$\tau_2 \in [0.53, 0.59]$ is followed by a period-halving
		route for $\tau_2 > 0.59$.}
	\label{fig:FigPoinc6}
\end{figure}

Within the alternation interval, the system shows sensitivity to the dual-delay coupling. At $\tau_2 = 0.53$, the dynamics present a stable period-1 orbit. However, a minimal parametric increase to $\tau_2 = 0.54$ shows a chaotic dynamics, verified by the aforementioned multi-diagnostic criteria and the trajectories shown in Figure~\ref{fig:Atractores_6}.a and b. This chaotic phase is suppressed almost instantly; at $\tau_2 = 0.55$, the orbits abruptly collapse onto a predictable period-2 limit cycle, corroborated by the recovery of a negative Lyapunov exponent. Nevertheless, when $\tau_2 = 0.56$ again chaos is found and generates the strange attractor shown in Figure~\ref{fig:Atractores_6}.b. This alternation between periodic cycles and chaos across increments in $\tau_2$ of order $10^{-2}$ or less suggests the signature of chaotic–periodic alternation behavior in the parameters.

\begin{figure}[t!]
	\centering
	\includegraphics[width=0.85\linewidth]{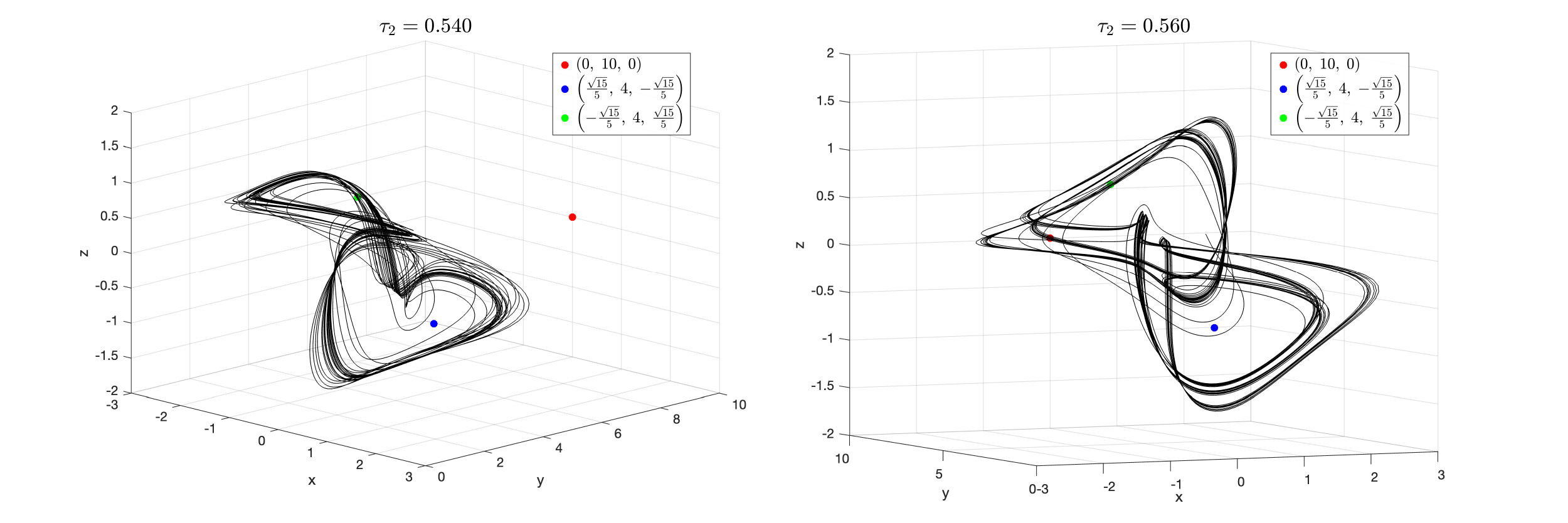}
%	\begin{subfigure}[b]{0.45\textwidth}
%		\includegraphics[width=\linewidth]{images/Atractor_6(1).png}
%		\caption{$\tau_2 = 0.54$}
%	\end{subfigure}\hfill
%	\begin{subfigure}[b]{0.45\textwidth}
%		\includegraphics[width=\linewidth]{images/Atractor_6(2).png}
%		\caption{$\tau_2 = 0.56$}
%	\end{subfigure}
	\caption{Strange attractors for $\tau_1 = 1.85$, $K = -0.098$, at the two
		chaotic values of $\tau_2$ within the alternation interval.}
	\label{fig:Atractores_6}
\end{figure}

Extending to $\tau_2 > 0.59$, the chaotic–periodic alternation ceases and the system returns to a period-halving route. The chaotic fluctuations are dampened, yielding to a period-2 orbit at $\tau_2 = 0.60$, which returns to a period-halving bifurcation to settle onto a fundamental period-1 limit cycle at $\tau_2 = 0.65$. This stabilization shows that while narrow bands of transmission delay can derive into volatility, sufficiently extended temporal lags eventually suppress the instability. As in previous cases, phase portraits are shown in Figure~\ref{fig:FigSim_6} along with their metrics in Table~\ref{tabla_6}.}

\begin{table}[t!]
	 
	\centering
	\caption{Dynamical regimes for $\tau_1 = 1.85$, $K = -0.098$, varying $\tau_2$.}
	\label{tabla_6}
	\begin{tabular}{r|cccccc}
		\toprule
		$\tau_2$ & $0.53$ & $0.54$ & $0.55$ & $0.56$ & $0.60$ & $0.65$ \\ 
		$\lambda_{\max}$ & $< 0$ & $> 0$ & $< 0$ & $> 0$ & $> 0$ & $\approx 0$\\
		Dynamics & Period-1 & Chaos & Period-2 & Chaos & Period-2 & Period-1\\
		\bottomrule
	\end{tabular}
\end{table}

\section{Discussion}
\label{sec:discussion}

{The numerical results presented in Section~\ref{sec:results} indicate that the complexity increases with the addition of heterogeneous time delays. In the baseline single-delay case, the system shows behaviors consistent with existing literature on the  delayed-feedback Chen  model~\citep{chen2008dynamics}, and models both cyclical and anti-cyclical macroeconomic interventions. Under these conditions, the system can show an  period-halving routes as well as bifurcations between periodic and chaotic regimes with chaotic–periodic alternation.

\begin{table}[ht!]
	\centering
	\caption{Comparison between the delayed feedback Chen financial system and the proposed heterogeneous delay model.}
	\label{tab:comparison}\small
	\begin{tabular}{p{0.17\textwidth}p{0.33\textwidth}p{0.43\textwidth}}
		\toprule
		\textbf{Feature} & \textbf{Chen (2008)} & \textbf{Present approach} \\ \midrule
		Model Base & Huang-Li-Ma-Chen (HLMC)  & Modified HLMC with Policy Feedback \citep{HuangLi1993,MaChen2001a} \\
		Delay Structure & Homogeneous ($\tau_1 = \tau_2 = \dots$)  & Heterogeneous ($\tau_1 \neq \tau_2$) \citep{Lakshmanan2010, Farmer1982} \\
		Feedback Type & Single-channel (independent)  & Pyragas-type on Interest Rate ($x$) \citep{Pyragas1992, Pyragas1995} \\
		Policy Parameter & Fixed/Positive feedback strength  & Bipolar Gain $K$ (Cyclic vs. Anti-cyclic) \citep{Pyragas1992,Holyst2000} \\
		Transmission & Instantaneous price response  & Delayed price index ($\tau_2$) \citep{Pyragas1992, Cesare2005} \\
		Routes to Chaos & Period-doubling cascades \citep{chen2008dynamics} & Period-Doubling, Period-Halving, and Incommensurable \citep{Flores2025,intermitencia} \\
		New Phenomenon & Attractor merging crises \citep{chen2008dynamics,Grebogi1983} & \textbf{Chaotic–periodic alternation} via delay coupling \citep{Flores2025,intermitencia} \\ \bottomrule
	\end{tabular}
\end{table}

The introduction of a second delay associated with the price index transmission also gives rise to incommensurable routes to chaos, diverging from the standard period-doubling cascades and attractor merging crises documented in earlier works, and then chaotic–periodic alternation appears as a  dynamical feature of the system. A summary contrasting the results of this work with those existing in the literature is provided in Table~\ref{tab:comparison}.}

%\subsection{Single-delay dynamics and the role of feedback polarity}

{ Under a single feedback delay $\tau_1$, the anti-cyclical case for values $K < 0$ and $\tau_1 = 1.85$, undergoes a classical period-doubling cascade. Throughout the presented results, it was shown how the numeric calculation of a positive maximum Lyapunov exponent \citep{Wolf1985} was not enough to determine the emergence of chaos, so more metrics have to be included to determine it, reinforcing its well-known caveat.

\subsection{Chaotic–Periodic Alternation}

{ The primary finding of this work is the existence of alternations between periodic and chaotic regimes when a second transmission delay $\tau_2$ is considered.} In the cyclical case, the system transitions repeatedly among chaos, incommensurable-period motion, and period-1 orbits as $\tau_2$ varies over $[0.9, 1.3]$. Transitions { occur on increments of $\tau_2$ as small as $\mathcal{O}(10^{-2})$}, and the sequence does not follow a monotone bifurcation pattern, as stable windows are repeatedly interrupted by chaotic regimes. In the anti-cyclical case, chaotic–periodic alternation is concentrated in $\tau_2 \in [0.53, 0.59]$, where transitions between period-1, chaos, and period-2 occur on { increments of $\tau_2$ of order $\mathcal{O}(10^{-2})$}, before the system settles into a period-halving route for $\tau_2 > 0.59$.

{To the best of the authors' knowledge, alternation of this type has not been documented in the double-delay integer-case variants of the delayed Chen  model~\citep{chen2008dynamics, ChenWC2008frac}. While the self-feedback delay $\tau_1$ governs the internal correction dynamics of the interest rate based on its own historical states, the cross-coupling transmission delay $\tau_2$ models the informational lag between the interest rate and the price index. When both delays are simultaneously active, they induce sudden changes in the form of transitions~\citep{Pomeau1980}. As a consequence of this coupling, the phase space is altered in ways that cannot be uncovered by studying each delay in isolation~\citep{Farmer1982, Lakshmanan2010}.}

\subsection{Economic interpretation}

The presented results can be useful for the analysis and design of financial policy interventions. The single-delay analysis establishes that Pyragas-type feedback policy that depends jointly on the feedback polarity $K$ and the implementation delay $\tau_1$, consistent with the literature on delayed feedback in economic models~\citep{Holyst2000, Cesare2005}, and confirms that the classification of a policy as cyclical or anti-cyclical is insufficient without simultaneous specification of the implementation lag.

However, the chaotic–periodic alternation result shows that when both $\tau_1$ and $K$ are selected to produce stable dynamics, the introduction of a secondary {transmission delay $\tau_2$ can induce} chaotic bursts. In practical terms, a stabilization policy tuned on the basis of the interest rate feedback channel alone may fail if the transmission time into prices differs from its nominal value by a small amount, but succeed in others, even when $\tau_2$ is present, that indicates that robust policy design requires explicit accounting for transmission lags within the system. Nevertheless, it must be noted that this extension serves as a low-dimensional abstraction of an economy, so the precise quantitative thresholds and bifurcation points reported in this work may not be directly transferable to empirical financial systems, but serve as a way to take into account more elements into it. 

Therefore, in this work we show how asymmetric multi-channel delays induce chaotic–periodic alternation. While real macroeconomic latencies exhibit continuous distributions, asynchronous reporting, and institutional friction, our low-dimensional model isolates this structural vulnerability through intentional simplification. %Extending this framework to higher-dimensional systems with empirical delay distributions represents an open direction for future research.

%{ While the heterogeneous delay framework introduced here uncovers novel pathways to instability, several structural boundaries must be acknowledged when contextualising these findings. Crucially, the proposed model serves as a stylised, low-dimensional abstraction of an economy. Consequently, the precise quantitative thresholds and bifurcation points reported in our numerical analysis are not directly transferable to empirical financial systems.

%The primary contribution of this work is qualitative and mechanistic rather than prescriptive; it isolates and proves the existence of chaotic intermittency driven purely by the interaction of asymmetric, multi-channel delays. In real-world macroeconomic environments, informational and policy latencies are rarely discrete or uniform. Instead, they are governed by continuously distributed lags, asynchronous reporting cycles, and evolving institutional friction. Therefore, while our low-dimensional model successfully captures the structural vulnerability of coupled systems, it represents an intentional simplification. Extending this analysis to higher-dimensional macroeconomic frameworks---particularly through the integration of empirically estimated delay distributions---remains an open problem for future research.}

\section{Conclusions}

In this work we show that complex dynamical transitions within an extended three-dimensional financial model can appear from the integration of an Pyragas-type feedback loop into the interest rate and a structural transmission delay $\tau_2$ within the price index. Cyclical feedback drives previously reported period-doubling cascades and quasi-periodic routes to chaos via incommensurable orbits. We also explore anti-cyclical policies that show inverse sequences where chaotic regimes collapse into stable periodic windows before re-destabilizing.

The main contribution shows that the simultaneous action of both delays induces abrupt alternations between chaotic bursts and ordered windows in the delayed system of Chen. As propagation lags induce both chaotic and periodic regimes, financial stabilization frameworks must account for multi-delay modeling that can lead to chaotic regimes. Extending this framework to higher-dimensional macro-dynamics, incorporating empirical lag estimation, stochastic noise, system interconnections, and partial information constraints, remains open for future research.

\section{Acknowledgements}
Authors thank UACM for its support via Project UACM-CCYT-2026-CON-12.

\section{Declaration of Generative AI and AI-assisted technologies in the writing process}
During the preparation of this work, the authors used Claude (Anthropic) to revise the language, grammar, and readability of the text. Following the use of this tool, all authors edited, and approved the final content as needed, and take full responsibility for the accuracy, integrity, and originality of the work.

\bibliographystyle{plainnat}
\bibliography{referencias}

\appendix
\section{Reproduction of the results of \cite{chen2008dynamics,MaChen2001a}}\label{Sec:ChenResults}
In this case, a sweeping for the delay in the Pyragas-type feedback over the interval $[1, 8]$ and a cyclical action with $K=0.1$ is done, and in Figure~\ref{fig:FigPoinc1} the bifurcation diagram obtained upon varying the implementation delay $\tau_1$ under a single-delay cyclical control configuration, reproducing the results of \cite{chen2008dynamics}.

\subsection{Fixed $K=0.1$, varying $\tau_1 \in [1,8]$} \label{subsec:cyclic1}
It can be seen that, in the first interval spanning $\tau_1 \in [1.5, 2]$, the system undergoes an abrupt transition from a periodic regime to a chaotic state, whereas the second interval, $\tau_1 \in [4, 8]$, is characterized by a progressive, supercritical period-doubling route to chaos. Within the first interval, specifically at $\tau_1 = 1.65$ and $\tau_1 = 1.75$, both the Poincare section and the phase portraits are consistent with stable, period-1 limit cycles, as seen in Fig. \ref{fig:FigSim_1}a-b., and within these two regimes Lyapunov exponents are $\lambda_{\max} \approx 0$. In addition, fundamental peaks in the Fourier power spectra and the regular, repeating oscillations of the trajectories confirm a non-chaotic state. Further, after a destabilization threshold at $\tau_1 = 1.85$, the system enters a chaotic regime marked by an positive Lyapunov profile ($\lambda_{\max} = 0.0621$), highly erratic aperiodic time series, and a broadband, continuous Fourier spectral density, as seen in \ref{fig:FigSim_1}.c., where it can be seen that it forms a strange attractor (see Figure~\ref{fig:Atractores_1}.a). 

As the parameter enters the second interval, a window of delay-induced re-stabilization starts at $\tau_1 = 6.40$, collapsing the trajectories back onto a single-loop period-1 limit cycle with a slightly negative exponent ($\lambda_{\max} = -0.00148$). This stable regime subsequently gives way to a period-doubling bifurcation at $\tau_1 = 6.70$, producing a double-looped period-2 orbit, as seen in \ref{fig:FigSim_1} in panels d) and e) respectively. Ultimately, at $\tau_1 = 7.02$, the system re-enters a chaotic regime characterized by a broadband Fourier spectrum, aperiodic solutions, and a positive exponent $\lambda_{\max} = 0.07334$, as seen in Fig. \ref{fig:FigSim_1}.f. This state generates a larger attractor (Fig.~\ref{fig:Atractores_1}.b). Phase portraits for each of these discrete regimes as well as spectral and maximum Lyapunov exponent are shown in Figure~\ref{fig:FigSim_1}, with their corresponding metrics summarized in Table~\ref{tabla_1}.

\begin{figure}[h!tbp]
	\centering
	\includegraphics[width=0.55\linewidth]{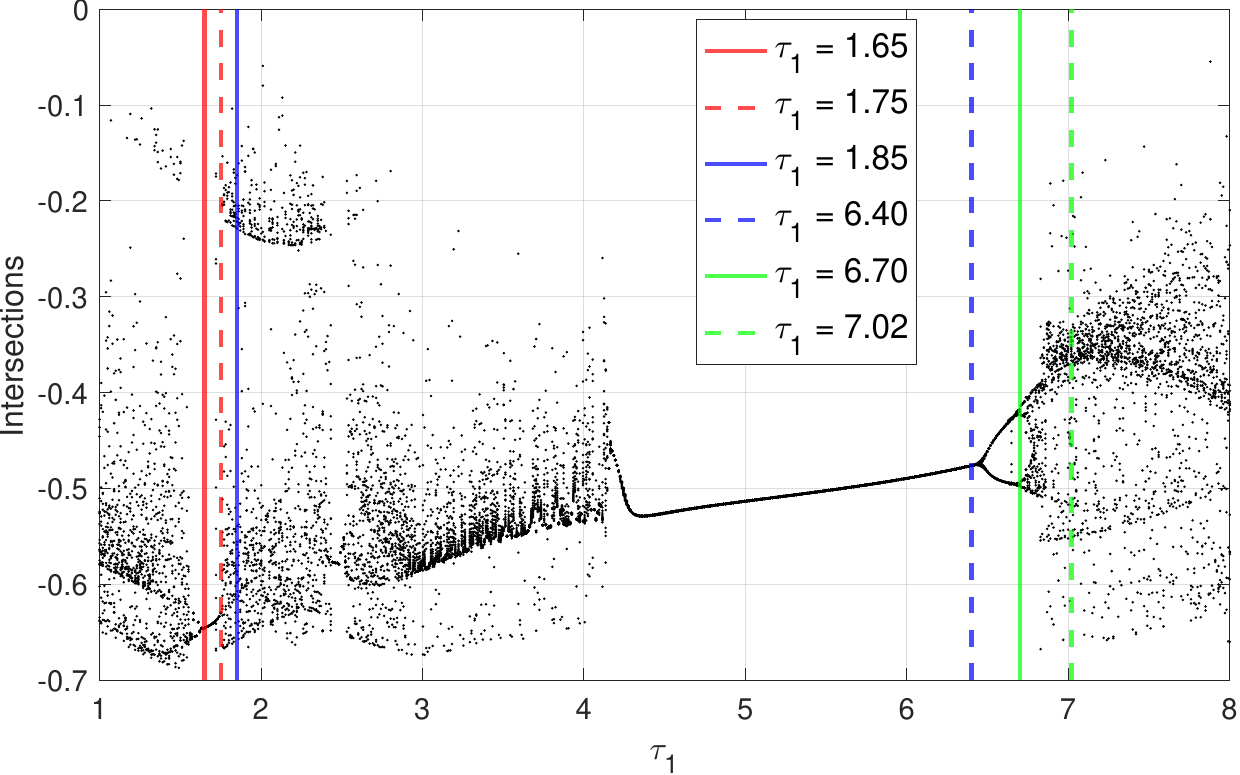}
	\caption{Poincaré diagram on the section $z = 0$ as a function
		of $\tau_1$, with $K = 0.1$ and $\tau_2 = 0$. An abrupt transition
		near $\tau_1 \approx 1.85$ { (blue solid vertical line)} and a period-doubling cascade for
		$\tau_1 \in [4,8]$ are visible.}
	\label{fig:FigPoinc1} %Simulación generada con /Users/marcos/Library/CloudStorage/Dropbox/paperChaosFinance/Simulaciones/Fig1_bifurcation_tau1.m
\end{figure}

\begin{figure}[t!]
	\centering
		\includegraphics[width=\linewidth]{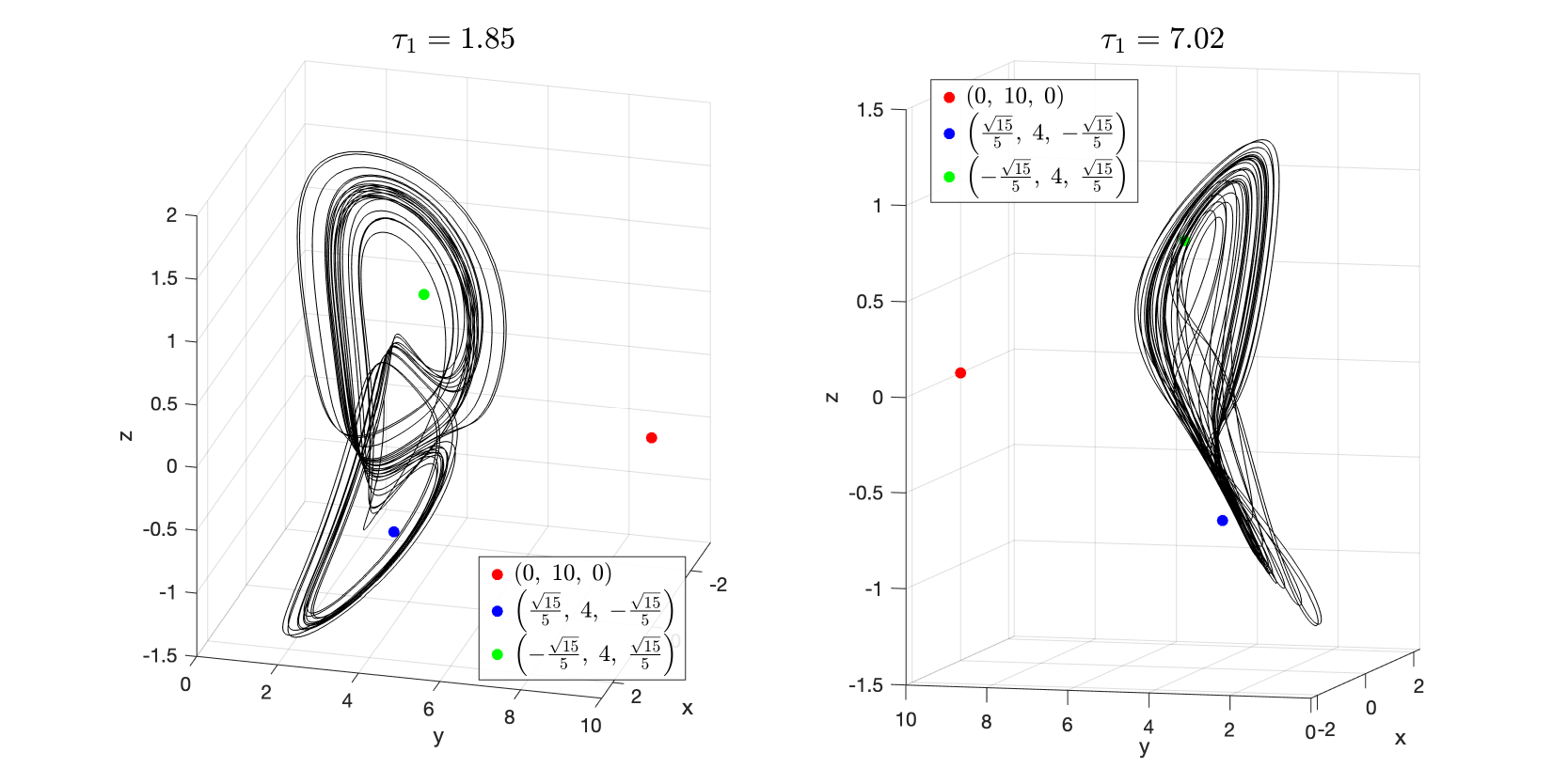}
	\caption{Strange attractors for $K = 0.1$, $\tau_2 = 0$, at the two chaotic parameter values identified in the bifurcation diagram. { Equilibrium points $E_0$ and $E_{\pm}$ are in red, blue and green, respectively}}
	\label{fig:Atractores_1}
\end{figure}

\begin{figure}[t!]
	\centering
	\begin{subfigure}[b]{0.48\textwidth}
		\includegraphics[width=\linewidth]{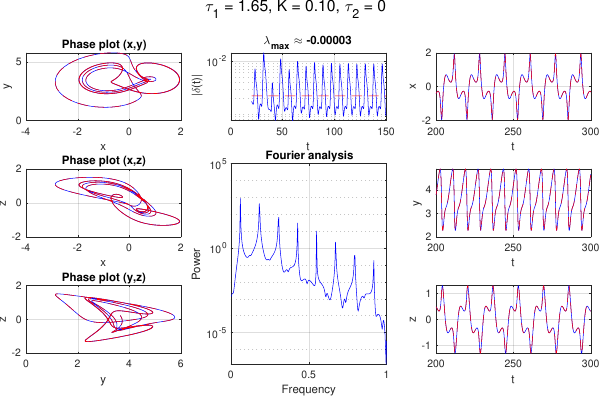}
		\caption{$\tau_1 = 1.65$ (period-1)}
	\end{subfigure}\hfill
	\begin{subfigure}[b]{0.48\textwidth}
		\includegraphics[width=\linewidth]{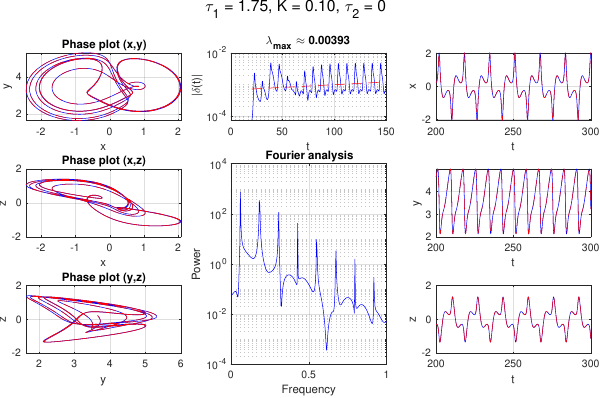}
		\caption{$\tau_1 = 1.75$ (period-1)}
	\end{subfigure}
	\vspace{0.3cm}
	\begin{subfigure}[b]{0.48\textwidth}
		\includegraphics[width=\linewidth]{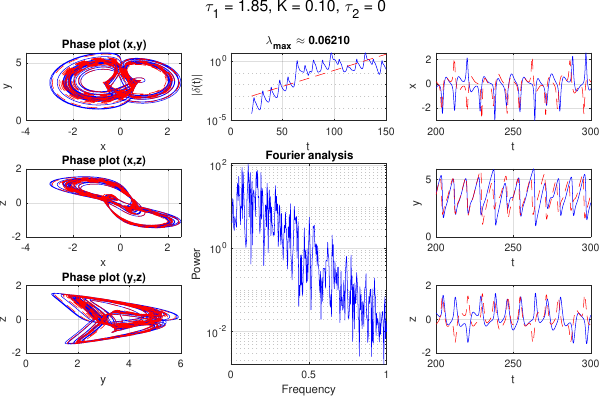}
		\caption{$\tau_1 = 1.85$ (chaos)}
	\end{subfigure}\hfill
	\begin{subfigure}[b]{0.48\textwidth}
		\includegraphics[width=\linewidth]{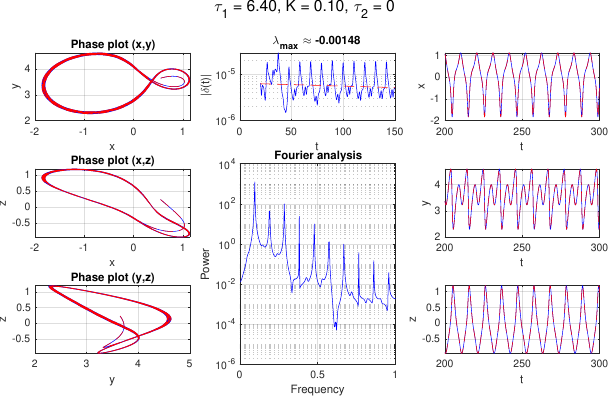}
		\caption{$\tau_1 = 6.40$ (period-1)}
	\end{subfigure}
	\vspace{0.3cm}
	\begin{subfigure}[b]{0.48\textwidth}
		\includegraphics[width=\linewidth]{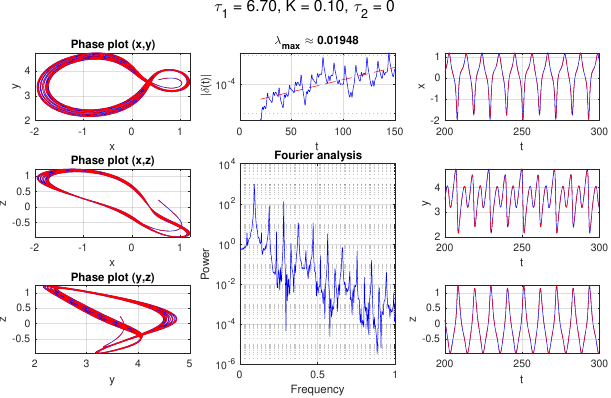}
		\caption{$\tau_1 = 6.70$ (period-2)}
	\end{subfigure}\hfill
	\begin{subfigure}[b]{0.48\textwidth}
		\includegraphics[width=\linewidth]{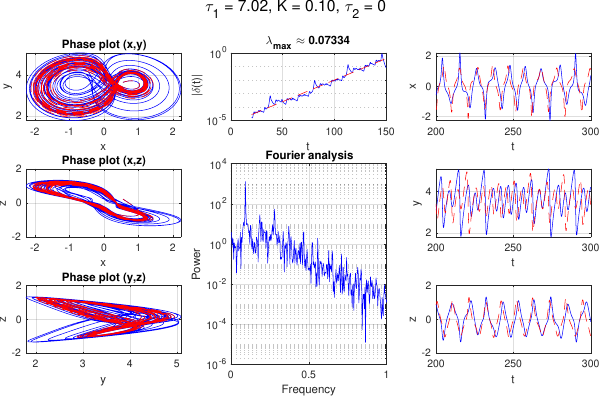}
		\caption{$\tau_1 = 7.02$ (chaos)}
	\end{subfigure}
	\caption{Phase portraits for $K = 0.1$, $\tau_2 = 0$, and selected values of $\tau_1$. Panels (a)--(b): period-1 motion. Panel (c): chaotic attractor ($\lambda_{\max} = 0.0621$). Panel (d): period-1 recovered. Panel (e): period-2 via period-doubling. Panel (f): chaos re-established. The time series depict two trajectories starting at slightly different initial conditions after the transient was omitted.} %Simulación y figuras generadas mediante /Users/marcos/Library/CloudStorage/Dropbox/paperChaosFinance/Simulaciones/Fig2_SimAndFigures_V2.m
	\label{fig:FigSim_1}
\end{figure}

%  __                                  _  
% (_   _   _ _|_ o  _  ._    |_|_ /|    ) 
% __) (/_ (_  |_ | (_) | |     | o | o /_ 
%                                         

\subsection{Fixed $\tau_1 = 6.40$, varying $K \in [0.25, 0.50]$}

Initiating from the fundamental period-1 regime localized at $(\tau_1, K) = (6.40, 0.1)$, the cyclical feedback gain $K$ is explored across the interval $K \in [0.25, 0.50]$, with the temporal delays held at $\tau_1 = 6.40$ and $\tau_2 = 0$. The resulting bifurcation mapping, which can be seen in Figure~\ref{fig:FigPoinc3}, illustrates a transition to deterministic chaos that avoids the well-known period-doubling cascade, progressing instead via a sequence of quasi-periodic regimes characterized by incommensurable-period orbits.

Phase-space projections for selected values are shown in Fig. \ref{fig:FigSim_3}. The corresponding Fourier power spectra across this $K$-values are filled by multiple discrete, incommensurable frequency peaks, which is a definitive spectral signature for quasi-periodic dynamics. Furthermore, the maximal Lyapunov exponent remains consistently and very nearly null, $\lambda_{\max} \approx 0$, which confirms the absence of exponential trajectory divergence and verifies that this incommensurable motion is non-chaotic, seen in in Fig. \ref{fig:FigSim_3}.a-e. However, notice that this marginal stability is abruptly lost at the critical threshold $K = 0.50$ in Fig. \ref{fig:FigSim_3}.f. A strange attractor forms as shown in Figure~\ref{fig:Atractor_2}. As in the previous case, selected values that show this route to chaos are shown in Figure~\ref{fig:FigSim_3}, and the related metrics metrics provided in Table~\ref{tabla_2}.

%FIGURAS - Dinámicas en  tau1=6.40, K in 0.25 - 0.50
\begin{figure}[t]
	\centering
	\includegraphics[width=0.75\linewidth]{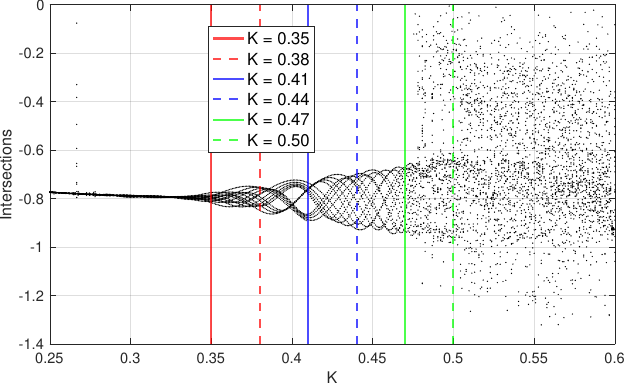}
	\caption{Bifurcation diagram on the Poincar\'{e} section $z = 0$ as a function
		of $K$, with $\tau_1 = 6.40$ and $\tau_2 = 0$. A transition through
		incommensurable-period dynamics to chaos near $K \approx 0.50$ is visible.}
	\label{fig:FigPoinc3}
\end{figure}

\begin{figure}[t!]
	\centering
	\begin{subfigure}[b]{0.48\textwidth}
		\includegraphics[width=\linewidth]{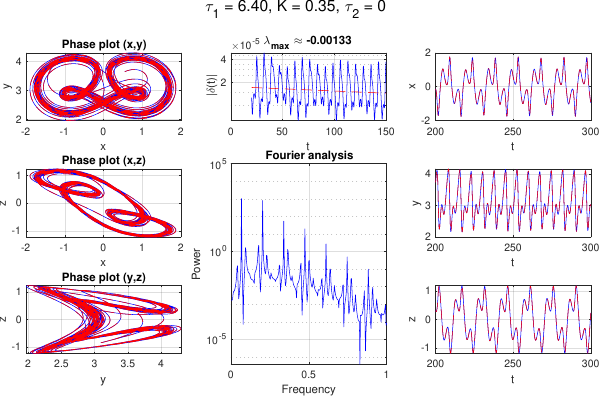}%FigSim_1(3).pdf
		\caption{$K = 0.35$}
	\end{subfigure}\hfill
	\begin{subfigure}[b]{0.48\textwidth}
		\includegraphics[width=\linewidth]{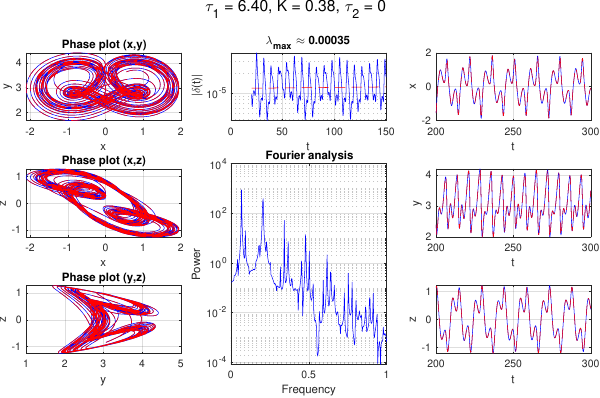}%FigSim_2(3).pdf
		\caption{$K = 0.38$}
	\end{subfigure}
	\vspace{0.3cm}
	\begin{subfigure}[b]{0.48\textwidth}
		\includegraphics[width=\linewidth]{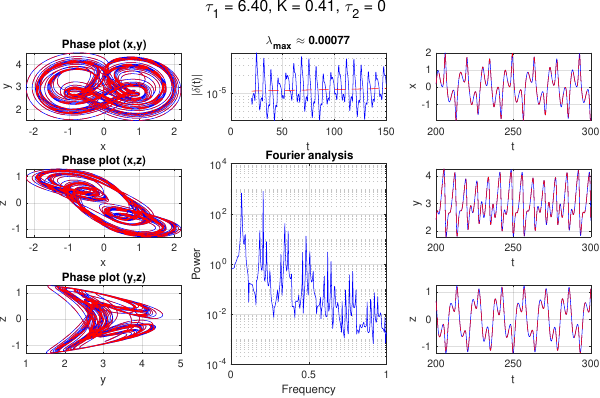}%FigSim_3(3).pdf
		\caption{$K = 0.41$}
	\end{subfigure}\hfill
	\begin{subfigure}[b]{0.48\textwidth}
		\includegraphics[width=\linewidth]{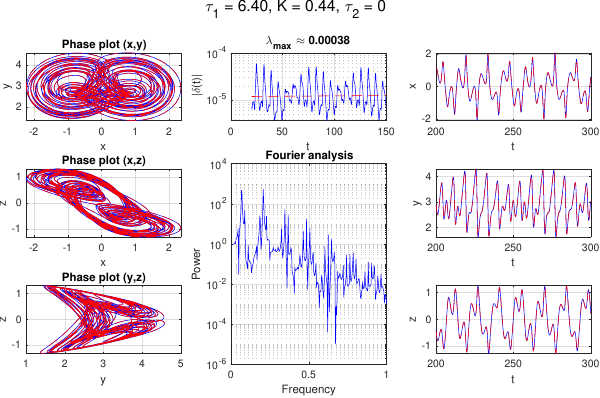}%FigSim_4(3).pdf
		\caption{$K = 0.44$}
	\end{subfigure}
	\vspace{0.3cm}
	\begin{subfigure}[b]{0.48\textwidth}
		\includegraphics[width=\linewidth]{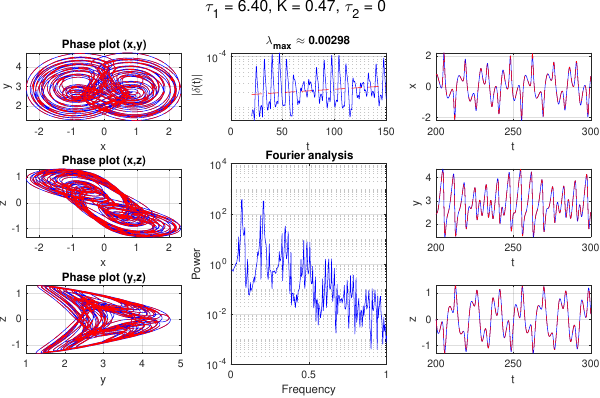}%FigSim_5(3).pdf
		\caption{$K = 0.47$}
	\end{subfigure}\hfill
	\begin{subfigure}[b]{0.48\textwidth}
		\includegraphics[width=\linewidth]{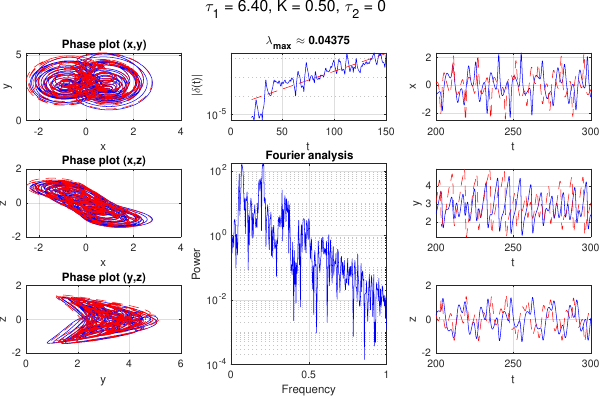}%FigSim_6(3).pdf
		\caption{$K = 0.50$ (chaos)}
	\end{subfigure}
	\caption{Phase portraits for $\tau_1 = 6.40$, $\tau_2 = 0$, and selected values
		of $K > 0$. Panels (a)--(e): incommensurable-period orbits
		($\lambda_{\max} \approx 0$). Panel (f): chaotic attractor.}
	\label{fig:FigSim_3}
\end{figure}

\begin{table}[t!]
	\centering
	 
	\caption{Dynamical regimes for $K = 0.1$, $\tau_2 = 0$ at selected $\tau_1$-values.}
	\label{tabla_1}
	\begin{tabular}{r|cccccc}
		\toprule
		$\tau_1$         & 1.65         & 1.75         & 1.85 & 6.40        & 6.70  & 7.02\\
		$\lambda_{\max}$ &  $\approx 0$ &  $\approx 0$ & $>0$ & $\approx 0$ & $> 0$ & $> 0$\\ %$0.08817$ for 1.85-value
		Dynamics & Period-1 & Period-1 & Chaos & Period-1 & Period-2 & Chaos\\
		\bottomrule
	\end{tabular}
\end{table}

%FIGURA - Strange attractor - tau1=6.40, K in 0.25 - 0.50
\begin{figure}[t]
	\centering
	\includegraphics[width=0.85\linewidth]{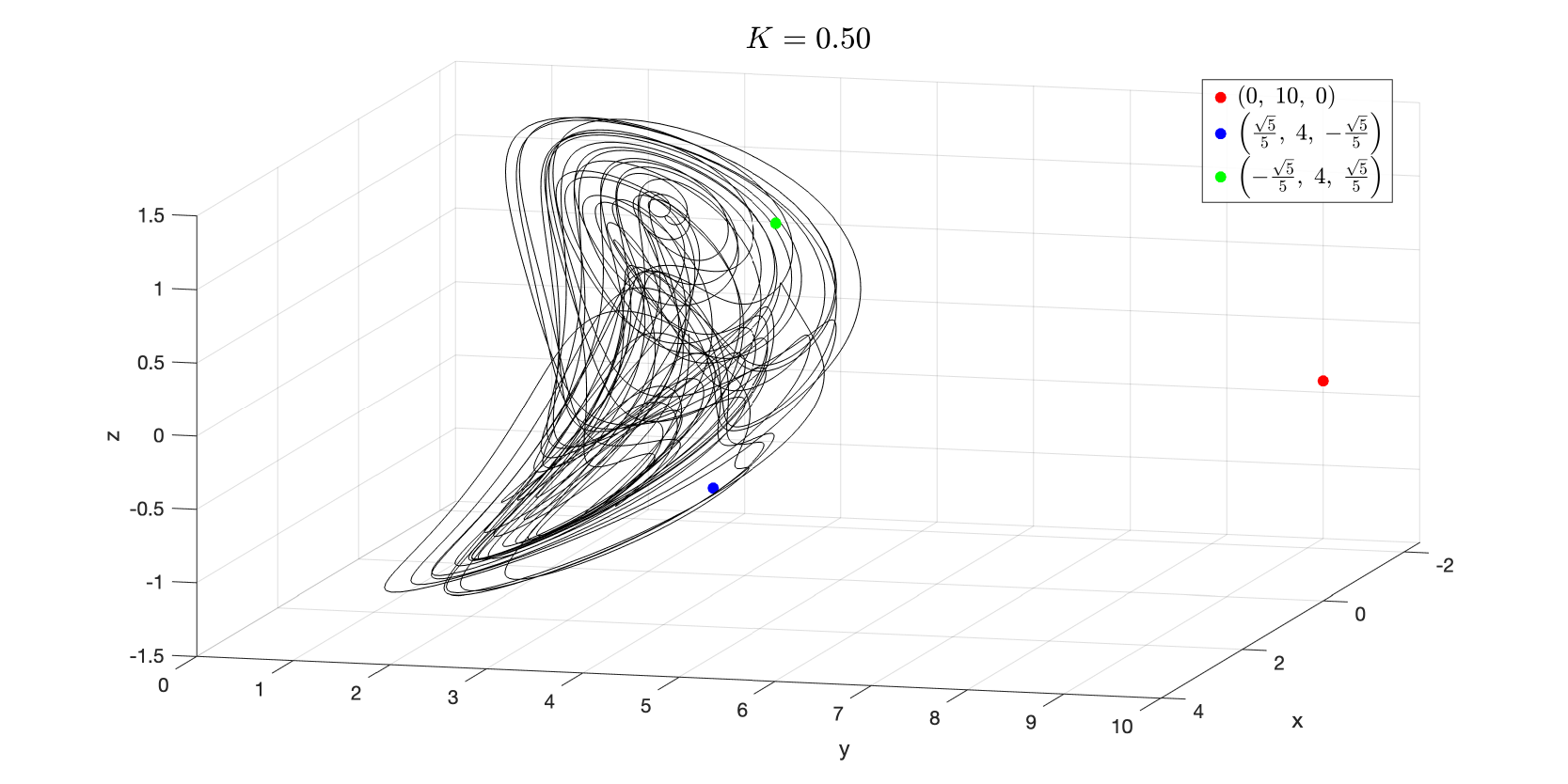}  %Atractor_2(1).png
	\caption{Strange attractor for $\tau_1 = 6.40$, $\tau_2 = 0$, $K = 0.50$.}
	\label{fig:Atractor_2}
\end{figure}

%TABLA - tau1=6.40, K in 0.25 - 0.50
\begin{table}[t!]	 
	\centering
	\caption{Dynamical regimes for $\tau_1 = 6.40$, $\tau_2 = 0$, varying $K > 0$.}
	\label{tabla_2}
	\begin{tabular}{r|ccc}
		\toprule
		$K$ & \{0.35\,, 0.38\,, 0.41\,, 0.44\,, 0.47\}  & 0.50 \\
		$\lambda_{\max}$ & $\approx 0$ & $> 0$ \\
		Dynamics & Incommensurable period & Chaos \\
		\bottomrule
	\end{tabular}
\end{table}

\end{document}